RESEARCH ARTICLE

SPACE SCIENCES

# Origin of uranium isotope variations in early solar nebula condensates

François L. H. Tissot,* Nicolas Dauphas, Lawrence Grossman

High-temperature condensates found in meteorites display uranium isotopic variations ($^{235}$U/$^{238}$U) that complicate dating the solar system's formation and whose origin remains mysterious. It is possible that these variations are due to the decay of the short-lived radionuclide $^{247}$Cm ($t_{1/2}$ = 15.6 My) into $^{235}$U, but they could also be due to uranium kinetic isotopic fractionation during condensation. We report uranium isotope measurements of meteoritic refractory inclusions that reveal excesses of $^{235}$U reaching ~+6‰ relative to average solar system composition, which can only be due to the decay of $^{247}$Cm. This allows us to constrain the $^{247}$Cm/$^{235}$U ratio at solar system formation to (7.0 ± 1.6) × 10$^{-5}$. This value provides new clues on the universality of the nucleosynthetic r-process of rapid neutron capture.

## INTRODUCTION

All elements beyond the iron peak (above ~70 atomic mass units) are the products of three main processes of stellar nucleosynthesis: the s- (slow neutron capture), r- (rapid neutron capture), and p-process (proton process) (1, 2). Unlike the s- and p-process, which are relatively well understood [neutron capture in asymptotic giant branch (AGB) stars for the s-process and photodisintegration of seed nuclei in supernovae for the p-process] (3, 4), little is known regarding the astrophysical conditions under which r-process nuclides are produced (5–8). The r-process label may comprise some diversity as it was suggested that up to three r-processes were responsible for producing light r-nuclides (A < 140), heavy r-nuclides (A > 140), and actinides (for example, U and Th). The existence of an "actinide" production site is motivated by two main observations: (i) the ages of some old stars inferred from their Th/Eu ratios are negative, meaning that they must have formed with a Th/Eu ratio that was higher than that relevant to the solar system (SS) (9), and (ii) the abundance of the short-lived radionuclide (SLR) $^{244}$Pu [$t_{1/2}$ = 79.3 million years (My)] in meteorites is low compared to another nominal r-process radionuclide, $^{182}$Hf (7). However, $^{244}$Pu has a long half-life and its stellar yield is uncertain (10), which makes it insensitive to the history of nucleosynthesis before SS formation and whether or not multiple r-process sites contributed to the synthesis of the actinides. In contrast, $^{247}$Cm, which decays into $^{235}$U, has a much shorter half-life ($t_{1/2}$ = 15.6 My) and would be very well suited to address this question. Unfortunately, its abundance in the early solar system (ESS) has been the subject of some debate that boils down to knowing whether the uranium isotope variations (that is, $^{235}$U/$^{238}$U ratio) measured in early-formed nebular condensates [calcium- and aluminum-rich inclusions (CAIs)] are due to the decay of $^{247}$Cm or isotopic fractionation during condensation. The two cannot be easily distinguished because U has only two long-lived isotopes and both mechanisms would produce similar correlations between light U isotope enrichments and U concentrations.

Recently, excesses in $^{235}$U of up to 3.5‰ were documented in four fine-grained CAIs (11). To demonstrate that those variations are due to $^{247}$Cm decay, one must show that the δ$^{235}$U values correlate with the Cm/U parent-to-daughter ratios. Because Cm has no long-lived isotopes, another element must be used as a proxy for Cm in isochron diagrams. On the basis of their near-identical valence states, ionic radii, and volatilities, the light rare earth elements (REEs) are thought to behave similarly to Pu and Cm during nebular processes (12–14). This conclusion is supported by the observed coherent behavior of Pu, Nd, and Sm in pyroxene/melt and phosphate/melt partitioning experiments (15) and during magmatic differentiation in achondrites (12, 14). Although Th has sometimes been used as a proxy for Cm (11, 16), coherent behavior for the Th-Cm and Th-Pu pairs is neither expected (13) nor observed (12), and the light REEs (for example, Nd or Sm) are therefore taken as more reliable proxies of Cm behavior during condensation.

The $^{235}$U excesses observed by Brennecka et al. (11) correlate with Nd/U ratios, which the authors interpreted as evidence of live $^{247}$Cm in the ESS at a level of ($^{247}$Cm/$^{235}$U)$_{ESS}$ = (1.1 to 2.4) × 10$^{-4}$. However, such $^{235}$U enrichments could also reflect mass-dependent fractionation during condensation of solid CAIs from nebular gas (17–19). Indeed, the kinetic theory of gases predicts that the lighter isotope ($^{235}$U) should condense faster than the heavier isotope ($^{238}$U), resulting in fractionations that can reach $\sqrt{238/235} \sim$ 6‰ for condensation of atomic U in a low-pressure gas. In both cases of $^{247}$Cm decay and fractionation during condensation, one would expect to find correlations between U isotopic variations and the degree of U depletion. It is therefore presently undecided which mechanism is responsible for the U isotope variations documented in CAIs.

Definitive evidence of live $^{247}$Cm therefore awaits the discovery of meteoritic material that is highly depleted in uranium and displays $^{235}$U excess outside of the ±6‰ window allowed by fractionation at condensation. Because the abundance of $^{247}$Cm in the ESS is expected to be low (10), such large excesses of $^{235}$U from $^{247}$Cm decay will only be resolvable in phases with $^{144}$Nd/$^{238}$U atomic ratios exceeding ~2100 (atomic ratios will be used hereafter). Therefore, numerous slabs of the Allende meteorite were examined, and 15 CAIs (12 fine-grained and 3 coarse-grained) were extracted and digested with acids. Many of these CAIs revealed group II REE patterns and large U depletions indicative of incomplete condensation of refractory lithophile elements (see the Supplementary Materials). A method was developed to measure accurately the U isotopic compositions of samples with low U contents (0.1 to 0.01 ng; see the Supplementary Materials).

Origins Lab, Department of the Geophysical Sciences, and Enrico Fermi Institute, University of Chicago, Chicago, IL 60637, USA.
*Corresponding author. E-mail: ftissot@uchicago.edu

## RESULTS

As in previous studies, most samples display low $^{144}$Nd/$^{238}$U ratios (that is, <900), and their $\delta^{235}$U values are within 6‰ of the bulk SS value at $\delta^{235}$U = +0.31‰ (relative to CRM-112a) (19, 20). These samples display a trend between $\delta^{235}$U and $^{144}$Nd/$^{238}$U similar to that described previously on similar samples (11) and which had been taken as evidence that $^{247}$Cm was alive in the ESS. The samples display significant scatter around the best-fit line that cannot be entirely explained by analytical uncertainty [the mean square weighted deviation (MSWD) of the regression for samples with $^{144}$Nd/$^{238}$U ratios lower than 900 is 46]. Even coarse-grained CAIs that are nondepleted in U (low $^{144}$Nd/$^{238}$U ratios) have $\delta^{235}$U that vary between −0.16 and +1.35‰, well outside of analytical uncertainty.

One sample (named Curious Marie) has an extremely high $^{144}$Nd/$^{238}$U ratio (~22,640) and a $^{235}$U excess of +58.9 ± 1.9‰ (equivalent to an absolute $^{238}$U/$^{235}$U ratio of 130.17 ± 0.27; Fig. 1 and Table 1). For comparison, the highest $^{144}$Nd/$^{238}$U ratio and $^{235}$U excess measured in CAIs prior to this work were 794 and +3.43‰ ($^{238}$U/$^{235}$U = 137.37), respectively (11). Considerable effort was expended to confirm this result (see the Supplementary Materials). The measurement was triplicated using different sample purification schemes and various measurement setups. The tests yielded $\delta^{235}$U values of +52.79 ± 14.91‰ after one purification step with $^{235}$U measured on Faraday, +59.12 ± 2.80‰ after two purifications with $^{235}$U on an electron multiplier, and +58.97 ± 2.72‰ after three purifications with $^{235}$U on an electron multiplier. Extensive testing was also done to ensure that no interferences or matrix effects were affecting the measurement by combining the matrix cut from Curious Marie with CRM-112a, purifying the U by column chemistry, and finding that the measured U isotopic composition is correct (that is, identical to pure CRM-112a after purification.

## DISCUSSION

### Evidence for $^{247}$Cm

The finding of such a large excess $^{235}$U in a normal CAI [in opposition to fractionated and unknown nuclear (FUN) CAIs, which is a group of refractory inclusions that display FUN effects; see the Supplementary Materials] can only be explained by decay of $^{247}$Cm into $^{235}$U. Below, we examine other processes that could potentially lead to U isotope variations in ESS materials and show that they all suffer from serious shortcomings:

(i) Isotopic fractionation during secondary processes (for example, aqueous alteration and/or redox processes), either on the meteorite parent body or on Earth, can impart large isotopic fractionation to light elements (such as Li). However, the degree of fractionation generally decreases as the mass of the element increases, and for U on Earth, the variations are limited to ~1.5 to 2‰ (21). Although this process could lead to some scatter in the data around the isochron, it cannot explain a +59‰ anomaly.

(ii) Large isotopic variations of nucleosynthetic origin have been documented for refractory elements in CAIs and are usually readily identified as departures from mass-dependent fractionation. For uranium, which has only two stable isotopes, it is impossible to distinguish between $^{247}$Cm decay and nucleosynthetic anomalies. Nevertheless, considering that anomalies on the order of a few permil in heavy elements (for example, Ba, Sm, and Nd) are only found in FUN CAIs (normal CAIs have more subdued anomalies of a few tenths of permil) and given that Curious Marie ($\delta^{238}$U ~ 59‰) is not a FUN CAI (see the Supplementary Materials), the large $^{235}$U excess found here cannot be reasonably ascribed to the presence of nucleosynthetic anomalies.

(iii) Finally, U isotopic fractionation could be the result of fractionation during evaporation/condensation processes. During condensation, the light isotope of U condenses faster, leading to large $^{235}$U depletion in the condensing gas and the instantaneous solid, but $^{235}$U excesses limited to +6‰ (see the Supplementary Materials). Similarly, during evaporation, the highest $^{235}$U excess predicted by the kinetic theory of gases is limited to ~+6‰. Consequently, evaporation/condensation processes cannot explain the +59‰ $^{235}$U excess observed in Curious Marie, which leaves $^{247}$Cm decay as the only possible explanation.

### Closure time in Curious Marie and $^{247}$Cm/$^{235}$U ratio in the ESS

The large excess of $^{235}$U found in the Curious Marie CAI is definitive evidence that $^{247}$Cm was alive in the ESS. An initial ($^{247}$Cm/$^{235}$U) ratio of ~5.6 × 10$^{-5}$ at the time of closure can be calculated using the slope of the isochron in Fig. 1. Even though the samples with low $^{144}$Nd/$^{238}$U ratios define a trend with $\delta^{235}$U, the slope of the isochron in Fig. 1 is mainly leveraged by Curious Marie ($^{144}$Nd/$^{238}$U ratio ~ 22,640). The initial $^{247}$Cm/$^{235}$U ratio of (5.6 ± 0.3) × 10$^{-5}$ thus corresponds to the time when this CAI acquired its high $^{144}$Nd/$^{238}$U ratio. Terrestrial alteration is ruled out because the Allende meteorite is an observed fall that did not experience much terrestrial weathering. The extreme uranium depletion in the Curious Marie CAI is thus most likely due to solar nebula condensation and/or nebular/parent body alteration. All fine-grained CAIs in this study display a typical group II

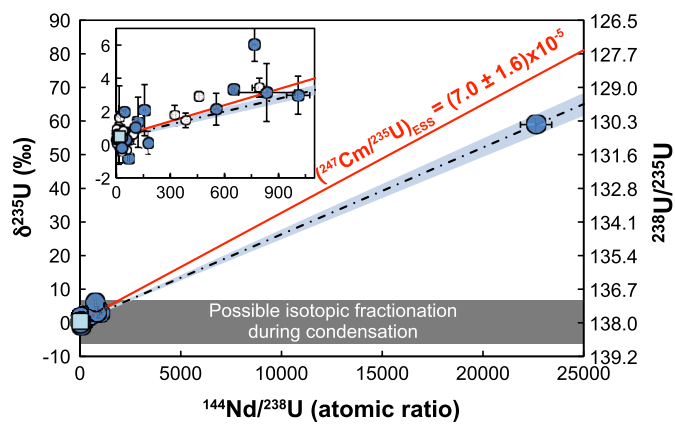

**Fig. 1. $\delta^{235}$U plotted as a function of the $^{144}$Nd/$^{238}$U atomic ratio in meteoritic samples.** Open circles, previous studies (11, 17–19, 33, 34); blue circles, Allende CAIs from this work; light-blue square, bulk Allende from this work]. The +59‰ $\delta^{235}$U value observed in the Curious Marie CAI is well outside the range of variations expected from fractionation during condensation (gray rectangle) and is thus interpreted as definitive evidence for live $^{247}$Cm in the ESS. The scatter in the data (for example, at very low Nd/U ratios) suggests that stable isotopic fractionation during evaporation/condensation also influenced the U isotopic composition of CAIs. The slope of the two-point isochron between Curious Marie and the rest of the samples translates into a $^{247}$Cm/$^{235}$U of (5.6 ± 0.3) × 10$^{-5}$ at the time of Nd/U fractionation in Curious Marie. Accounting for a possible delay between this fractionation event (possibly related to the extensive alteration of this CAI) and the formation of the SS of 5 ± 5 My, the inferred $^{247}$Cm/$^{235}$U at SS formation is (7.0 ± 1.6) × 10$^{-5}$ (red line).

Table 1. Type, REE pattern, mass, U content, $^{144}$Nd/$^{238}$U atomic ratio, and U isotopic composition of the samples analyzed in this study. $n$ is the number of replicate analyses for each sample (starting from digested sample). Nd/U ratios and isotopic compositions are corrected for blank contributions (see table S2). $\delta^{235}$U = [($^{235}$U/$^{238}$U)$_{sample}$/($^{235}$U/$^{238}$U)$_{CRM-112a}$ − 1] × 10$^3$. Error bars are 95% confidence intervals. Nd/U ratio was calculated using the U concentration from the double-spike technique, and Nd concentration from the standard addition technique (see the Supplementary Materials).

| Average $^{144}$Nd/$^{238}$U ratios and U isotopic compositions of CAIs | | | | | | | | | |
|---|---|---|---|---|---|---|---|---|---|
| Sample | Type | REE pattern | Mass (mg) | U (ng) | n | $^{144}$Nd/$^{238}$U | ± | $\delta^{235}$U (‰) blk corr. | ± |
| Allende | CV3 | | 1016 | 21.9 | 2 | 20.2 | 2.5 | 0.49 | 0.16 |
| FG-1 | Fine-gr. CAI | Group II | 35.2 | 0.77 | 2 | 156.7 | 11.6 | 2.08 | 1.54 |
| Curious Marie | Fine-gr. CAI | Group II | 717.9 | 0.37 | 3 | 22,640 | 780 | 58.93 | 2.08 |
| CG-1 | Coarse-gr. CAI | Group I | 48.4 | 0.76 | 2 | 120.8 | 8.8 | 1.35 | 1.52 |
| FG-2 | Fine-gr. CAI | Group II | 50.4 | 1.53 | 2 | 173.9 | 6.5 | 0.12 | 0.69 |
| FG-3 | Fine-gr. CAI | Group II | 98.3 | 4.05 | 2 | 104.8 | 4.5 | 1.01 | 0.50 |
| FG-4 | Fine-gr. CAI | Group II | 349.7 | 15.7 | 2 | 66.8 | 4.7 | −0.83 | 0.25 |
| FG-5 | Fine-gr. CAI | Group II | 48.7 | 2.82 | 2 | 54.2 | 3.9 | 0.30 | 0.59 |
| FG-6 | Fine-gr. CAI | Group II | 54.9 | 0.35 | 2 | 1009 | 64 | 2.98 | 1.14 |
| FG-7 | Fine-gr. CAI | Group II | 61.4 | 0.98 | 2 | 555 | 16 | 2.13 | 0.96 |
| FG-8 | Fine-gr. CAI | Group II | 377.3 | 8.50 | 2 | 653 | 10 | 3.32 | 0.21 |
| FG-9 | Fine-gr. CAI | Group II | 330.7 | 23.9 | 2 | 44.2 | 3.1 | 1.95 | 0.14 |
| FG-10 | Fine-gr. CAI | Group II | 15.61 | 0.24 | 2 | 834 | 156 | 3.14 | 1.75 |
| CG-2 | Coarse-gr. CAI | Group V | 200.4 | 23.0 | 2 | 31.5 | 2.2 | 0.43 | 0.14 |
| FG-11 | Fine-gr. CAI | Group II | 68.6 | 0.83 | 2 | 768 | 16 | 6.03 | 1.02 |
| TS32 | Coarse-gr. CAI | Group V | 41.8 | 4.89 | 2 | 32 | 2.4 | −0.16 | 0.30 |

REE pattern, thought to represent a snapshot in time and space of the condensation sequence (22, 23). The most refractory REEs (heavy REEs except Tm and Yb) are depleted in these CAIs because they were sequestered in ultrarefractory dust such as perovskite or hibonite, whereas the more volatile REEs (Eu and Yb) and uranium are also depleted because they stayed in the gas phase when those CAIs formed. In most samples, U and Yb present similar levels of depletion relative to solar composition and the abundance of other refractory lithophile elements (Fig. 2), indicating that those two elements have similar behaviors during evaporation/condensation processes under solar nebula conditions. However, in Curious Marie, the U/Nd ratio is 1000 times lower than solar composition, whereas the Yb/Nd ratio is only depleted by a factor of 50. If U and Yb have similar behaviors during condensation, one would expect the U/Nd ratio to be 50 times lower than solar, not 1000 times as is observed. Compared to other fine-grained CAIs analyzed, Curious Marie is peculiar because it is extremely altered, which is manifested by the extensive replacement of high-T phases by low-T alteration products such as nepheline and sodalite [its Na$_2$O is 15 weight percent (wt %) when other CAIs are all lower than 6.7 wt %; table S1]. Such alteration may have mobilized U, producing a 20-fold U depletion on top of the 50-fold depletion associated with condensation.

Some of these alteration products could have formed (i) in the nebula, (ii) on an earlier generation of water-rich asteroid, or (iii) during aqueous alteration on Allende itself [for example, Ross et al. (24) and Russell and MacPherson (25) and references therein]. Regardless of the location, dating of aqueous alteration products on meteorite parent bodies with extinct radionuclides $^{36}$Cl ($t_{1/2}$ = 0.301 My), $^{26}$Al ($t_{1/2}$ = 0.717 My), $^{53}$Mn ($t_{1/2}$ = 3.74 My), or $^{129}$I ($t_{1/2}$ = 15.7 My) suggests that it took place no later than 10 My after SS formation. This is a conservative upper limit because secondary alteration phases in fine-grained inclusions may have formed very early in the nebula. The half-life of $^{247}$Cm is 15.6 My, meaning that a time span of ~5 ± 5 My between SS formation and CAI alteration (and possible U depletion) translates into a correction for the initial $^{247}$Cm/$^{235}$U ratio of 25 ± 25% (the uncertainty on this factor takes into account the uncertainty on the closure age). Our present best estimate of the initial SS $^{247}$Cm/$^{235}$U ratio is thus (7.0 ± 1.6) × 10$^{−5}$, which is equivalent to ($^{247}$Cm/$^{238}$U)$_{ESS}$ = (2.2 ± 0.5) × 10$^{−5}$ and ($^{247}$Cm/$^{232}$Th)$_{ESS}$ = (9.7 ± 2.2) × 10$^{−6}$. This value is in agreement with the $^{247}$Cm/$^{235}$U ratio of (1.1 to 2.4) × 10$^{−4}$ obtained by Brennecka et al. (11) based on CAI measurements and an upper limit of ~4 × 10$^{−3}$ inferred from earlier meteoritic measurements (26). It is also in line with the lower estimate derived from modeling of galactic chemical evolution (GCE) (10), which predicts an initial ratio of (5.0 ± 2.5) × 10$^{−5}$.

### Fractionation of U isotopes during evaporation/condensation

Some of the samples with low $^{144}$Nd/$^{238}$U ratios show significant scatter around the isochron (see inset of Fig. 1), outside of analytical uncertainties. In particular, samples with $^{144}$Nd/$^{238}$U ratio as low as ~30 span a range of $\delta^{235}$U values of 3.5‰, a feature that led previous workers (*17*, *18*) to question the conclusion of Brennecka et al. (*11*) that $^{247}$Cm was responsible for U isotope variations. This scatter is most likely due to isotopic fractionation during condensation, suggesting that better isochronous behavior could be obtained if such fractionation could be corrected.

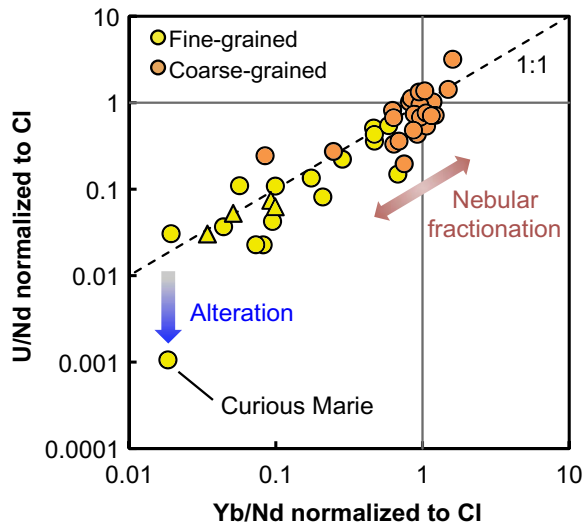

**Fig. 2. Correlation between U and Yb abundances relative to solar composition and the abundance of Nd (a refractory lithophile element).** Circles denote data from different studies (*35*–*37*) and this study, and triangles denote data from Brennecka et al. (*11*). This correlation indicates that U and Yb have similar behaviors during evaporation/condensation processes under solar nebula conditions. The Curious Marie CAI plots off the 1:1 line, with a CI-normalized U/Nd ratio 20 times lower than that of Yb/Nd. This CAI is extremely altered (see main text and table S1), and such alteration may have mobilized U, producing a 20-fold U depletion on top of the 50-fold depletion associated with condensation.

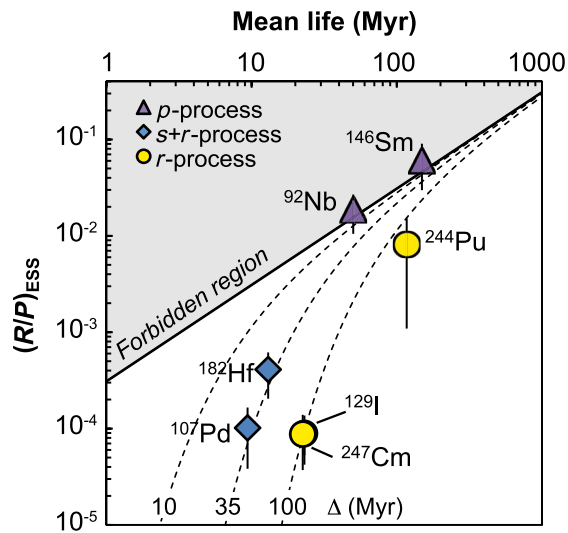

**Fig. 3. Meteoritic abundance ratios of extinct radionuclides to stable nuclides produced by the same process (for example, $^{129}$I/$^{127}$I$^r$), normalized to stellar production ratios versus mean lives ($\tau = t_{1/2}/\ln 2$).** The superscript "*r*" refers to the *r*-process component of the cosmic abundance [obtained after subtracting the *s*-process contribution from solar abundances (*3*)]. When the normalizing isotope is not stable (for example, X/$^{238}$U or X/$^{232}$Th), the *R/P* ratio [that is, ($N_{SLR}/N_{Stable}$)/($P_{SLR}/P_{Stable}$)] is corrected for the decay of the long-lived isotope by multiplication by the *N/P* ratio of the normalizing isotope in the ESS [0.71 for $^{238}$U and 0.89 for $^{232}$Th, values from Nittler and Dauphas (*10*)]. Triangles denote *p*-process isotopes, diamonds denote *r*-process nuclides with possible large *s*-process contributions, and circles denote *r*-process isotopes. The data are compared to model steady-state abundances in the ISM, using the model of Dauphas et al. (*28*) with $k = 1.7$ and a presolar age of the galaxy of 8.7 Gy. Dotted curves show model abundances assuming free-decay intervals of 10, 35, and 100 My, respectively. The abundances of all *r*-process nuclides can be explained by a single *r*-process environment that was active throughout the history of the galaxy and from which the solar system parent material was isolated about 100 My before SS formation, provided that $^{182}$Hf and $^{107}$Pd in meteorites originate from the *s*-process (*6*, *31*). See the Supplementary Materials for source data.

### Implications for the *r*-process

In addition to the dating implications that large $^{235}$U/$^{238}$U variations have on the Pb-Pb ages of the CAIs (*11*, *21*), the existence of $^{247}$Cm in the ESS has implications for the nucleosynthesis of *r*-process elements. The simplest GCE model that successfully reproduces the metallicity distribution of G-dwarfs requires infall of gas onto the galactic disk (*10*, *27*, *28*). In such a model, the abundance ratio of an SLR normalized to a stable nuclide is

$$\frac{N_{SLR}}{N_{Stable}} = \frac{P_{SLR}}{P_{Stable}} \cdot (k+1)\tau/T \quad (1)$$

where *N* is the abundance, *P* is the production ratio, $\tau$ is the mean life of the SLR ($\tau = 1/\lambda$), *T* is the age of the galaxy at the time when the ratio $N_{SLR}/N_{Stable}$ is to be calculated, and *k* is a constant that distinguishes closed-box ($k = 0$) versus infall models (typically, $k = 2 \pm 1$) (*8*, *27*–*29*). If *T* is taken as the presolar age of the galaxy, $T^*$ (= $T_G - T_{SS}$ = 8.7 Gy), then the ratio $N_{SLR}/N_{Stable}$ in the interstellar medium (ISM) at the time of isolation of the protosolar molecular cloud from fresh nucleosynthetic input can be calculated. This value can be compared to the $N_{SLR}/N_{Stable}$ ratio in the ESS as obtained from meteoritic measurements. The difference between the two values is often interpreted as a "free-decay interval" ($\Delta$) between the last nucleosynthetic event that produced the SLR and the formation of the SS

$$\left(\frac{N_{SLR}}{N_{Stable}}\right)_{ESS} = \left(\frac{N_{SLR}}{N_{Stable}}\right)_{ISM} \cdot e^{\left(-\frac{\Delta}{\tau}\right)} \quad (2)$$

At present, the meteoritic abundances of only three short-lived *r*-nuclides ($^{129}$I, $^{182}$Hf, and $^{244}$Pu) have been estimated, yielding three different $\Delta$ values (100 ± 7 My, ~35 My, and 158 ± 85 My, respectively) (*7*). The limit of our understanding of the *r*-process is illustrated by the fact that there have been as many *r*-processes proposed as short-lived *r*-nuclides investigated.

Using the value of ($^{247}$Cm/$^{232}$Th)$_{ESS}$ = (9.7 ± 2.2) × 10$^{-6}$ obtained in this study and the open nonlinear GCE model of Dauphas et al. (*28*) with $k = 1.7$, we obtained a free-decay interval of $\Delta = 98 \pm 14$ My. This value is in agreement with the $\Delta$ value of ~100 My derived from $^{129}$I and $^{244}$Pu but is much longer than the value of ~35 My obtained from $^{107}$Pd and $^{182}$Hf (Fig. 3). If $^{107}$Pd and $^{182}$Hf are indeed pure *r*-process isotopes, then a multiplicity of *r*-process environments is needed to explain the inconsistent $\Delta$ values (*5*, *7*). This is evident when looking at Fig. 3: $^{107}$Pd and $^{129}$I have similar ($N_{SLR}/N_{Stable}$)/($P_{SLR}/P_{Stable}$) ratios but different half-lives, making it impossible for both isotopes to be produced in a single event/process, no matter what model is used for the evolution of the SLR abundances in the giant molecular cloud parental to the SS [for example, free-decay interval versus three-phase mixing ISM (8, 30)]. However, recent nucleosynthetic models have reconsidered the origin of $^{107}$Pd and $^{182}$Hf and find a significant s-process contribution (70 to 80%) for both isotopes (6, 31). In such a framework, the initial abundance of all r-process SLR in the ESS can be explained by a single r-process making environment, which last injected material into the protosolar molecular cloud ~100 My before SS formation.

Although some low metallicity halo stars point to the existence of at least three distinct r-processes, these stars formed from a gas that had been enriched in r-process nuclides by synthesis in very low metallicity stars formed early in the history of the galaxy. Meteorite evidence suggests that such r-process multiplicity may only be relevant to exotic conditions that prevailed in the earliest generation of stars in the history of the galaxy and that a single r-process may still be relevant to long-term models of the chemical evolution of the galaxy.

## MATERIALS AND METHODS

Twelve fine-grained and three coarse-grained CAIs were selected for this study (fig. S1). One of the coarse-grained CAIs (TS32) was obtained in powder form directly and was described in a previous publication (32). All other CAIs were identified in Allende slabs and extracted with clean stainless steel dental tools before digestion by acids. A small chip of each CAI (all but TS32) was extracted using clean stainless steel dental tools under a stereoscopic zoom microscope, and mounted in epoxy for characterization.

All samples were mapped using a JEOL JSM-5800LV scanning electron microscope (SEM). Images of a selected field of view of each CAI, secondary electron, backscattered electron, and false-color RGB (Mg/Ca/Al) are shown in figs. S2 to S15. The REE patterns of the samples were determined using a Quadrupole LA-ICPMS (laser ablation inductively coupled plasma mass spectrometry) at the Field Museum. Diagnostic group II REE patterns were identified in all 12 fine-grained CAIs, indicating that the Nd/U ratio (a proxy for the Cm/U ratio) of these samples was high and that these samples were therefore well suited to search for $^{235}$U excesses coming from $^{247}$Cm decay.

After digestion but before column chemistry, a small aliquot of each sample (~2 to 3%) was used to determine U and REE concentrations. The samples were spiked with IRMM-3636 U double spike ($^{233}$U-$^{236}$U) and processed through U/TEVA resin (Eichrom) column chromatography, using Optima-grade and sub-boiling double-distilled acids.

Isotopic analyses were performed on a ThermoFinnigan Neptune MC-ICPMS at the Origins Lab (University of Chicago), equipped with an OnToolBooster 150 Jet Pump (Pfeiffer) and using Jet sample cones and X skimmer cones. An Aridus II desolvating nebulizer was used for sample introduction, and the measurements were done using a static cup configuration [see (21)]. When a ~+55‰ $^{235}$U excess was discovered in the Curious Marie CAI, a new measurement setup was developed that allowed characterization of the U isotopic composition of 0.1 to 0.01 ng of U with a precision of ±2 to 3‰ (see the Supplementary Materials). Extensive testing was done to ensure that no systematic bias was affecting the data, the details of which can be found in the Supplementary Materials.

## SUPPLEMENTARY MATERIALS

Supplementary material for this article is available at http://advances.sciencemag.org/cgi/content/full/2/3/e1501400/DC1
Materials and Methods
Curious Marie: not a FUN CAI.
The $^{247}$Cm-$^{235}$U chronometer and the initial abundance of $^{247}$Cm in the ESS.
GCE model
Fig. S1. Photos of typical fine-grained and coarse-grained CAIs.
Figs. S2 to S15. Secondary electron, backscattered electron, and false-color RGB maps of all samples.
Fig. S16. REE and U-Th abundance patterns of all 12 fine-grained CAIs analyzed in this study.
Fig. S17. Results of the standard addition measurements conducted on the Curious Marie CAI.
Fig. S18. U blank from new U/TEVA resin as a function of the volume of 0.05 M HCl passed through the column.
Fig. S19. Results of precision tests of U isotopic measurements done using various instrumental setups.
Fig. S20. Comparison of the $\delta^{235}$U determined on the CAIs using 80% (x axis) and 20% (y axis) of the sample.
Fig. S21. Flowchart of the tests conducted on the Curious Marie CAI.
Fig. S22. Evolution of the U isotopic composition of the gas, instantaneous solid, and cumulative solid, as a function of the fraction of U condensed.
Table S1. Results of SEM analysis on small chips of CAIs mounted in epoxy (wt %).
Table S2. Summary of U isotopic compositions and concentrations of CAIs and geostandards.
Table S3. Specifics of U isotopic measurements on MC-ICPMS for low U samples.
Table S4. Compilation of chemistry blanks and effect on U "stable" isotope ratio measurements.
Table S5. Summary of the Ti data obtained on geostandards and the Curious Marie CAI.
Table S6. Production ratios of selected SLRs produced by the s-, r-, and p-process and present in the ESS, normalized to a stable isotope produced in the same or similar nucleosynthetic process.
Table S7. Compilation of Cm/U isochron data and free-decay interval data from experimental and theoretical studies.
Table S8. Selected extinct radionuclides produced by the s-, r-, and p-process.
References (38–78)

**Acknowledgments:** We thank P. Heck, J. Holstein, and the Robert A. Pritzker Center for Meteoritics and Polar Studies at the Field Museum for providing some of the CAI samples; I. Steele for help in operating the SEM at the University of Chicago; and L. Dussubieux for help in operating the LA-ICPMS at the Field Museum. We are grateful to A. M. Davis and B. S. Meyer for discussions. **Funding:** This work was supported by grants from NASA (Laboratory Analysis of Returned Samples, NNX14AK09G; Cosmochemistry, OJ-30381-0036A and NNX15AJ25G) and NSF (Petrology and Geochemistry, EAR144495; Cooperative Studies of the Earth's Deep Interior, EAR150259) to N.D. and a NASA grant (Cosmochemistry, NNX13AE73G) to L.G. **Author contributions:** N.D. and F.L.H.T. designed the research; F.L.H.T. and L.G. selected the samples; F.L.H.T. performed the research; F.L.H.T., N.D., and L.G. wrote the paper. **Competing interests:** The authors declare that they have no competing interests. **Data and materials availability:** All data needed to evaluate the conclusions in the paper are present in the paper and/or the Supplementary Materials. Additional data are available from the authors upon request. This is Origins Lab contribution number 91.

Submitted 7 October 2015
Accepted 12 January 2016
Published 4 March 2016
10.1126/sciadv.1501400

Citation: F. L. H. Tissot, N. Dauphas, L. Grossman, Origin of uranium isotope variations in early solar nebula condensates. *Sci. Adv.* **2**, e1501400 (2016).

**The PDF file includes:**

**Materials and Methods:**

*Sample selection*

Given that the concentration of U in fine-grained CAIs is low (~20 ppb), large samples are needed for precise isotopic analysis. After visual inspection of numerous (>200) slabs of Allende, the largest fine-grained CAIs were selected for this study along with three coarse-grained CAIs. Fine-grained CAIs are readily identifiable from their texture (μm-sized grains), color (from white/grey to purple, with most being pink due to FeO in spinel) and irregular shapes. By comparison, coarse-grained CAIs are mostly white, with much coarser grains (mm- sized) than fine-grained CAIs (fig. S1). Most coarse-grained CAIs originated as molten droplets. The chemical compositions of fine-grained CAIs are those expected from partial condensation of refractory elements from solar composition gas (*38, 39*). Since group II REE patterns (*37, 40*) and significant U depletions are characteristic of fine-grained CAIs, 12 fine-grained CAIs were selected for this study to search for $^{235}$U excesses unambiguously resulting from $^{247}$Cm decay: *i.e.*, $^{235}$U excesses larger than the 6‰ expected to arise from evaporation/condensation processes (according to the kinetic theory of gases).

The coarse-grained CAIs were taken to serve as reference samples because the effect of $^{247}$Cm decay in coarse-grained CAIs is expected to be much smaller than in fine-grained CAIs. Indeed, the weight Nd/U ratios (a proxy for the Cm/U ratio) in fine-grained and coarse-grained CAIs are typically around 1079 and 110, respectively [average from (*11, 35-37, 41*)], which is equivalent to atomic $^{144}$Nd/$^{238}$U ratios of 297 and 39, respectively (when normalized to CI chondrites the ratios are 17.8 and 1.8, respectively). This one order of magnitude difference in the Cm/U ratio will directly translate into a one order of magnitude difference in the size of the expected isotopic anomalies. In a δ$^{235}$U *vs.* Cm/U plot (isochron plot), the fine-grained CAIs will define the slope of the isochron, and the coarse-grained CAIs, the intercept at the origin.

*Sample characterization*

As the collection name of each sample can be rather long and cumbersome, a shorter name was adopted for each CAI. The shortened and complete names of the CAIs can be found in table S1.

A small chip of each CAI was extracted using cleaned stainless steel dental tools (Hu-Friedy®), under a stereoscopic zoom microscope, and mounted in epoxy for characterization. All samples were mapped using a JEOL JSM-5800LV SEM (figs. S2 to S15, table S1). Typical grain size was about 10 μm for the fine-grained CAIs, which were mostly composed of spinel, pyroxene, sodalite and nepheline, with small and varying amounts of grossular and melilite, and some rare olivine, in good agreement with (*41*). The coarse-grained CAIs were mostly made of melilite, pyroxene, anorthite, and spinel, with a small amount of grossular and perovskite, in agreement with (*36*).

Following SEM characterization, all samples were then analyzed by LA-ICPMS using a Varian quadrupole ICP-MS connected to a New Wave UP213 laser ablation system at the Field Museum (Chicago). Ablation spots were 80 μm in diameter, the firing frequency was 15 Hz and energy densities were about 16.71 J/cm$^2$. Sample measurements were bracketed by measurements of NIST-614 glass, which contains sixty-one trace elements at a concentration of about 1 ppm (*42*). For each CAI, up to 4 spots were analyzed for ~ 35 sec and averaged in order to obtain an "average" CAI composition (see ablation spot locations on figs. S2 to S15). All twelve fine-grained CAIs were found to present a group II REE pattern (fig. S16), while coarse-grained CAI CG-2 showed an essentially flat pattern at about 18 times CI (group V REE pattern), and coarse-grained CAI CG-1 showed a flat pattern for the light REEs (9× CI) with a positive anomaly in Eu (15× CI) (group I REE pattern).

The LA-ICPMS study indicated that many of the CAIs surveyed were depleted in uranium, so those CAIs were suitable to search for the past presence of $^{247}$Cm. Powders of each sample were then extracted using cleaned dental tools under a stereoscopic microscope, and collected on weighing paper before being transferred into triple cleaned Teflon beakers. Great care was brought into collecting only the CAI and any meteoritic matrix accidentally extracted was removed from the collected fraction using cleaned tweezers. Sample masses ranged from 15 to 760 mg (table S2). Note that one of the coarse-grained CAIs (TS32) used in this study was characterized in an earlier study (*32*) and was obtained directly in powder form for this work. The REE pattern of TS32 is essentially flat at enrichment of 21× CI (group V REE pattern).

### Sample digestion

All samples were fully digested using Optima grade acids with two one-week attacks in HF/HNO$_3$ 3:1 (+ drops of HClO$_4$) followed by two one-week attacks in HCl/HNO$_3$ 2:1 on hot plates at 160°C. During each acid attack, the beakers were placed in a sonicator for ~1h. After the four acid attacks, the samples were taken back in concentrated HNO$_3$ and put back on hot plates for 4 days, before dilution to 3 M HNO$_3$. All samples were transferred into cleaned centrifuge tubes and centrifuged for 5 min at ~1500 rpm. No residue was visible.

At this stage, one fifth of the solution was saved for future work, leaving 80 % of the sample available for characterization of the U isotopic composition. A small aliquot (between 0.5 and 4 %) of the solution was sampled for concentration analyses on the Q-ICPMS at the Field Museum.

### Nd and U concentrations

Given that the values $^{144}$Nd/$^{238}$U ratios are of major importance in defining the slope of the isochron shown on Fig. 1, significant effort was expended to measure precisely and accurately the Nd and U concentrations in the CAIs.

For Nd, the concentrations were measured by Quadrupole inductively coupled plasma mass spectrometry (Q-ICPMS) for all CAIs, and by standard addition using Alfa Aesar Nd single solution (1000 µg/g, Lot# 801149A) for some selected samples (*Curious Marie*, FG-2, FG-3, FG-6, FG-7, FG-8, FG-10 and FG-11). Both sets of values agree with each other within less than 7 %. For Q-ICPMS measurements, the typical uncertainty is ± 10%. The uncertainties on the standard addition measurement are the 95 % confidence interval (typically ~2 to 6 %) and were calculated using the R statistical software. For instance, fig. S17 shows the results of the standard addition carried on *Curious Marie*, with the 95 % CI envelop.

For U, the concentrations were obtained on the MC-ICPMS from (i) the double-spike data reduction (*21*) when samples were spiked, and (ii) sample-standard bracketing when samples were not spiked. Each sample was measured twice (see table S2) and the U concentrations obtained in the two analyses typically agree within 5 to 20 %. The double-spike values are more precise and accurate than the bracketing values. For each U isotopic measurement, a value of the $^{144}Nd/^{238}U$ ratio of the sample can be calculated (with its corresponding 95 % confidence interval value). The final $^{144}Nd/^{238}U$ ratio used for a given sample is the weighted average of the individual $^{144}Nd/^{238}U$ ratios obtained for each sample (see table S2 and Table 1).

*Double spiking and uranium purification*

The samples were spiked (after digestion) using the commercially available IRMM-3636 double spike (50.45 % $^{233}U$ and 49.51 % $^{236}U$, (*43*)). Since the two spike isotopes ($^{233}U$, $^{236}U$) do not occur naturally and the spike is very pure (low $^{235}U$ and $^{238}U$ content), optimization of the spike to sample ratio is not critical. A minimum amount of spike, however, needs to be used to achieve high precision measurement of the spike isotopes while minimizing amplifier noise, as well as limiting the abundance sensitivity effect of $^{238}U$ on the $^{236}U$. To fulfill these requirements, previous studies used a $U_{Spike}/U_{Sample}$ ratio of ~3 % [*e.g.*, (*44*)]. As underspiking would be more detrimental than overspiking, all samples were spiked, aiming for a $U_{Spike}/U_{Sample}$ ratio of 6 %. The actual $U_{Spike}/U_{Sample}$ ratio was measured at ~ 6 % for all but one sample (*Curious Marie*, whose spiking level was only 0.5 %, table S2). This sample is the one with the largest digested mass (740 mg) and the underestimation in the U content is likely due to matrix effect during concentration measurements by Q-ICPMS (the concentration calculated from the measured U isotopic ratio is immune from such effects and is accurate).

To ensure full equilibration of the spike with the sample (*45*), the samples were dried completely after spike addition, taken back into concentrated $HNO_3$ and diluted to 3 M $HNO_3$. No residues were visible after this new digestion step.

U separation and purification was done using U/Teva resin, following the method described in (*21*). Due to the low U content of some of the samples, extensive column cleaning was done using 40 ml of 0.05 M HCl, removing essentially all U initially bound to the resin (down to the pg level, fig S18). This extensive cleaning is important as the amount of U bound to new U/Teva resin cartridges was typically 0.02 ng, which represents up to 70 % of the total U content of some of the samples measured in this study (table S2). The rest of the chemistry went as follows: (1) conditioning with 10 ml of 3M $HNO_3$, (2) sample loading in 3 M $HNO_3$, (3) matrix elution with 30 ml of 3 M $HNO_3$, (4) resin conversion to HCl with 5 ml of 10 M HCl, (5) Th elution in 12 ml of 5 M HCl and (6) U elution with 32 ml of 0.05 M HCl.

The U cut was then dried completely, and taken back into 200 μl of 1:1 $HNO_3/H_2O_2$. This step oxidizes the organics leached from the resin into the U cut. The samples were then dried again completely and taken back into 0.1 mL of concentrated $HNO_3$, before being evaporated to near-dryness and being finally taken back in 0.3 M $HNO_3$. At this stage, the samples were ready for mass spectrometric analysis. Some samples showed either particles or the solution did not wet the Teflon beakers like pure 0.3 M $HNO_3$ (*i.e.*, sticky aspect). In those rare instances, the last drying sequence (hydrogen peroxide/concentrated nitric/dilute nitric) was repeated until particles were dissolved and the solution behaved like pure 0.3 M $HNO_3$.

*Mass spectrometry*

Though the sensitivity of the MC-ICPMS is excellent for U isotopes (1.4 to 1.6 V/ppb), the total amount of U contained in fine-grained CAIs is small (typically 1 to 5 ng). Sensitivity tests were done to determine the precision of the measurements on the $^{235}U/^{238}U$ ratio as a function of the $^{238}U$ intensity measured, using a standard solution spiked with IRMM-3636 double spike (fig. S19, left panel). To resolve $\delta^{235}U$ anomalies on the order of 3 ‰, the $^{238}U$ signal must be at least 2 V (which translates into a minimum sample concentration of ~1.25 ppb). Therefore, after U purification, a 5 μl aliquot was taken from each sample (~2.5 % of the sample) to determine the quantity of U recovered. The sample volume (and thus its U concentration) was adjusted in order to optimize the precision achievable for each CAI. The samples with the least U (0.3 to 3.7 ng) were measured in 0.25 to 0.50 mL, allowing for only a single measurement with no chance of error.

Details of the cup configuration used for the measurements are given in table S3. To not waste any sample, the take-up time was set to zero and the cycles leading to the intensity plateau were manually removed during data reduction. Baseline measurements were done before each analytical session and at least daily, by defocussing the beam for 100 cycles of 1.05 second.

When it was found that *Curious Marie* had a ~+50 ‰ anomaly in $\delta^{235}U$, a new measurement setup was designed in which sufficient precision would be achieved to re-measure this unique sample while using only 10 % of the original sample solution (*i.e.*, half of the aliquot saved before sample spiking, containing about 0.04 ng of U). This was achieved by measuring $^{235}U$ on the SEM and $^{238}U$ in a Faraday cup with a $10^{11}$ Ω resistor. In this setup, the sample is not spiked, as the signal of the spiked isotopes would be too low (< 6 mV) for the double spike technique to be useful. Instead, the measurements are done using sample-standard bracketing. Results of precision tests using this configuration, with 2.097 seconds integration time and 100 cycles, are shown in fig. S19 (right panel). From this figure, it is clear that higher precision can be obtained for very low U concentrations by measuring $^{235}U$ on the SEM rather than on a $10^{12}$ Ω resistor.

*Data quality control*

Given the very low amount of U contained in the *Curious Marie* CAI and the unprecedented magnitude of the $^{235}$U excess found in this sample, considerable effort was spent to make sure that no analytical bias could account for any of the anomalies found in this sample and others.

### 1. General tests

Abundance sensitivity (the fact that the tail of the $^{238}$U peak is still present at mass 236, 235 and 234) can potentially affect the measurement of the $^{235}$U/$^{238}$U ratio. The typical contributions of the $^{238}$U tails on $^{236}$U, $^{235}$U and $^{234}$U, represent 0.6 ppm, 0.25 ppm and 0.1 ppm of the $^{238}$U signal, respectively (*21*). Not correcting for this effect only affects the $^{235}$U/$^{238}$U ratio by 0.1 ‰. Furthermore, the samples and bracketing standards are spiked to similar levels and their concentrations are matched within at least 10 %, thus the actual effect of the abundance sensitivity correction is smaller than 0.1 ‰. Therefore, abundance sensitivity issues cannot explain the U isotopic variations documented here.

Hydride formation could also be an issue. The ratio $^{238}$UH$^+$/$^{238}$U was regularly monitored and found to be on the order of 5×10$^{-7}$, in agreement with literature data (*21, 44, 46*). At this level, hydrides have no influence on the measurements. The hydride ratio would need to be on the order of 10$^{-3}$ to produce an anomaly of 2 ‰. Since the hydride formation ratio is so low, the possibility that $^{232}$ThH$^+$ interferences on $^{233}$U could affect the double spike correction is also ruled out. Again, the samples and bracketing standards are spiked at the same level, which would anyhow mitigate the influence of uranium hydrides on isotopic analyses.

To test if mismatch in acid molarity between the sample and standard could affect the measurements (*47, 48*), a 1 ppb CRM-112a solution (*i.e.*, the minimum concentration at which the measurements were performed with $^{235}$U on a Faraday cup) was spiked with IRMM-3636 (U$_{Spike}$/U$_{Sample}$ = 6 %) and several aliquots were dried and taken back in 0.3 M, 0.5 M, 1 M, 1.5 M, 2 M, and 3 M HNO$_3$. All these solutions were measured against a 0.3 M solution and the analyses yielded normal results within the 2 ‰ reproducibility of the measurements at this concentration.

As one sample was underspiked (U$_{Spike}$/U$_{Sample}$ = 0.5 %), a series of more than 50 standard measurements were performed with a U$_{Spike}$/U$_{Sample}$ ratio of 0.5 % to (i) check that no systematic bias was introduced by underspiking, and (ii) assess the reproducibility of a measurement carried out at such a spiking level. The data did not show any bias within uncertainty, and the standard value obtained was +1.7 ± 14.7 ‰ relative to the CRM-112a certified value.

The effect of the chemistry blank was also investigated. The $^{235}$U/$^{238}$U ratios of six chemistry blanks measured over a four month period were compiled and their effect on a sample measured at 1 ppb (*i.e.*, 1 to 1.5 V on $^{238}$U depending on the sensitivity of the instrument) was calculated (table S4). The effect was found to be between -0.9 and 2.8 ‰ at most.

Since the samples are dried with $H_2O_2$ right before analysis, a series of tests were done in which 17 to 54 mg of $H_2O_2$ was added to 0.25 ml of a 1 ppb standard solution (spiked at 6 %) before analysis. No effect on the $^{235}U/^{238}U$ ratio was found.

## 2. Assessment of data accuracy

Small amount of matrix elements can still be present after chemical purification and affect the accuracy of the measurement (*47, 48*). Two tests were performed to make sure that no interference or matrix effect influenced the results.

(a) The four samples with the most U (8 to 20 ng) were analyzed twice. Half the sample was analyzed after one purification step, while the other half was subjected to a second step of column chemistry before analysis. For all samples, both results are identical within error bars (~0.1 ‰) (table S2).

(b) For all samples (but the above four), the matrix cuts retrieved from the column chromatography were recombined and doped with a quantity of U standard CRM-112a similar to the amount of U initially present in the sample. These "doped" samples were passed through chemistry once before analysis, and for all but one, the $\delta^{235}U$ measured was indistinguishable from the CRM-112a value ($\delta^{235}U_{CRM-112a} = 0$ ‰, by definition), indicating a lack of interference/matrix effect for these samples.

One "doped" sample presented a large $^{235}U$ deficit (- 42 ‰): CAI FG-10. The same test was repeated for this sample, and once again, a $^{235}U$ deficit was found (-18 ‰). These values are in complete disagreement with each other, as well as with the measurement of the actual sample done using 80 % of the sample solution, which was measured at +0.48 ± 2.27 ‰. Since the U added to the sample for the doping test is the CRM-112a standard, and since the sample is double spiked and the yield above 90 %, the only plausible explanation is that some interference on $^{238}U$ is responsible for the $^{235}U$ deficit observed in the "doped" sample. This hypothesis was tested by looking at the $^{234}U/^{235}U$ ratio measured during the doping test, which was indistinguishable from the one of the standard, therefore indicating that the anomalous $^{235}U/^{238}U$ ratio is due to an interference on $^{238}U$. Such interference is possibly coming from insufficiently cleaned beakers/centrifuge tubes used in this test, as less care was expended for this particular task. The $\delta^{235}U$ value of FG-10 obtained using 80 % of the sample solution after one column chemistry is thus non-reliable and neither considered in the discussion nor plotted in the figures.

Two geostandards (BCR-2, BHVO-2) and a sample of Allende were prepared along with the CAIs to ensure that no systematic biases were introduced by the sample preparation. The mass of sample used was calculated to give ~20 ng of U for each geostandards, corresponding to the highest level of U contained in the CAIs used in this study. The $^{235}U/^{238}U$ ratios determined for the three geostandards are in excellent agreement with literature data [(*21*) and references therein].

Before spiking and chemistry, the samples were split 80 / 20, with the first aliquot being used for U analysis and the second being saved for future work. Some samples that contained high U contents were replicated by splitting the 80 % aliquot into two (see above). For the others, it was decided to use the 20 % aliquot to get a duplicate measurement. The 20 % aliquot was used and passed through the chemistry three times to make sure that no matrix elements remained after purification. The sample duplicates were within measurement error of the original values for all samples (fig. S20), except for FG-3 and TS32, for which the two replicates (with $^{235}$U measured on Faraday in both replicate measurements) are just outside of uncertainties of each other. These two samples have low $^{144}$Nd/$^{238}$U (105 and 33, respectively) and do not weigh in the determination of the slope of the isochron, and thus do not influence the main result of this study, *i.e.*, the initial abundance of $^{247}$Cm in the ESS.

   3. The Curious Marie CAI

Figure S21 summarizes the tests done on *Curious Marie*. In addition to the "doping" test described above, the $^{235}$U/$^{238}$U ratio of this sample was measured three times:

   (i) first using 80 % of the original sample solution spiked with IRMM-3636 and passed through the column chemistry once. All isotopes were measured using *Faraday* cups (Faraday setup in table S3)

   (ii) then using ~10 % of the original sample solution, non-spiked, and passed through column chemistry twice. The low abundance isotope $^{235}$U was measured using the Secondary Electron Multiplier (*SEM*) collector (SEM setup in table S3)

   (ii) and finally using ~5 % of the original sample solution, non-spiked, and passed through column chemistry three times. The low abundance isotope $^{235}$U was measured using the SEM setup.

Despite important differences in the protocols of chemical purification and isotopic analysis, all these three measurements yielded identical results (within uncertainties). This demonstrates that the large $^{235}$U excess measured in *Curious Marie* is real. This conclusion is strengthened by the fact that after three column passes, the acid chemistry blank was high enough (~0.01 ng of U) for its $^{235}$U/$^{238}$U ratio to be measured using the *SEM* setup. The blank composition was found to be indistinguishable from that of natural U (0.8 ± 9.2 ‰).

*Curious Marie*: not a FUN CAI

Though nucleosynthetic anomalies in centimeter-sized objects are usually very limited, some CAIs with *Fractionation and Unknown Nuclear* effects (the so-called FUN CAIs) display very large mass-dependent isotope effects and isotopic anomalies [*e.g.*, HAL, EK1-4-1 or TE, see, respectively (*49-51*)]. If *Curious Marie* were a FUN CAI, the large $^{235}$U excess found in this CAI could potentially be of nucleosynthetic origin. *Curious Marie*, however, does not present any of the characteristics of FUN CAIs. Indeed, FUN CAIs are all coarse-grained Type B, or compact Type A CAIs, whose oxygen isotopic compositions, when plotted in a three-isotope diagram, fall away from the CAI mass fractionation line. They are further recognized thanks to the large $\delta^{25}$Mg mass fractionation effects (> ~+10 ‰), large Ti mass-dependent effects (~ 10 ‰/amu), large $^{50}$Ti nucleosynthetic anomalies, and/or a deficit in $\delta^{26}$Mg* [for a review about FUN CAIs see (*52*)]. In contrast, *Curious Marie* is a fine-grained CAI, whose oxygen isotopic composition falls on the CCAM line (*53*), with resolvable $^{26}$Mg excess, negative $\delta^{25}$Mg values [-9 to -19 ‰, see (*53*)], a very minor $^{50}$Ti anomaly (~0.9 ‰, table S5), and no strong Ti mass-dependent fractionation (~ -0.28 ‰/amu, table S5). The negative $\delta^{25}$Mg observed in Curious Marie seems to be a common feature amongst fine-grained CAIs [see for instance EGG-5,

$\delta^{25}$Mg = -9 ‰; B29, $\delta^{25}$Mg = -9.6 ‰ and BG82DI, $\delta^{25}$Mg = -12.20 ‰; in (*54*) and references therein], whereas coarse-grained CAIs have $\delta^{25}$Mg values of about +5 to +10‰, and FUN CAIs have $\delta^{25}$Mg values in excess of +10 ‰ [see (*52, 54*)].

Altogether, the petrology, lack of Ti mass-dependent and independent effects, the negative $\delta^{25}$Mg fractionation and excess $^{26}$Mg, and the normal-like oxygen isotopic composition clearly indicate that *Curious Marie* is not a FUN CAI and, therefore, a nucleosynthetic origin for the $^{235}$U excess found in this CAI is extremely unlikely.

## The $^{247}$Cm-$^{235}$U chronometer and the initial abundance of $^{247}$Cm in the ESS

The Cm/U chronometer is based on the decay of $^{247}$Cm into $^{235}$U, which decays ultimately into $^{207}$Pb (*26, 33, 55*). Below we summarize the main features of this chronometer. Calculation of the abundance of $^{235}$U at any given time can be obtained by resolution of a differential equation of the kind commonly used in U-series studies (*56*)

$$\frac{d\,^{235}U}{dt} = -\lambda_{235}\,^{235}U + \lambda_{247}\,^{247}Cm \qquad (S.1)$$

Solving this equation gives the present day cosmic abundance of $^{235}$U, noted $^{235}$U$_P$, as

$$^{235}U_P = {}^{247}Cm_0 \left(\frac{\lambda_{247}}{\lambda_{235}-\lambda_{247}}\right)\left(e^{-\lambda_{247}\cdot t} - e^{-\lambda_{235}\cdot t}\right) + {}^{235}U_0 e^{-\lambda_{235}\cdot t} \qquad (S.2)$$

where the subscript P stands for "present", the subscript 0 denotes the initial number of atoms at closure of the system, and $\lambda_i$ the decay constant of the isotope i. The first term on the right hand side of the equation represents the contribution of $^{247}$Cm to the total $^{235}$U content, while the second term represents the contribution of the initial amount of $^{235}$U in the sample at the start of the solar system. Equation S.2 can be divided by the present cosmic abundance of $^{238}$U, $^{238}U_P = {}^{238}U_0 \cdot e^{-\lambda_{238}\cdot t}$, to obtain:

$$\left(\frac{^{235}U}{^{238}U}\right)_{smp,P} = \frac{^{247}Cm_{smp,0}}{^{238}U_{smp,p}}\left(\frac{\lambda_{247}}{\lambda_{235}-\lambda_{247}}\right)\left(e^{-\lambda_{247}\cdot t} - e^{-\lambda_{235}\cdot t}\right) + \left(\frac{^{235}U}{^{238}U}\right)_{smp,0}\frac{e^{-\lambda_{235}\cdot t}}{e^{-\lambda_{238}\cdot t}} \qquad (S.3)$$

where the subscript "smp" stands for "sample". Cm having no stable isotopes, a stable isotope proxy needs to be used in order to build an isochron. Nd, Sm, or Th are three elements that are thought to behave like Cm during nebular processes (*13*). Following most previous studies, we use $^{144}$Nd as proxy for Cm. We thus obtain:

$$\left(\frac{^{235}U}{^{238}U}\right)_{smp,P} = \left(\frac{^{247}Cm}{^{144}Nd}\right)_{smp,0}\left(\frac{\lambda_{247}}{\lambda_{235}-\lambda_{247}}\right)\left(e^{-\lambda_{247}\cdot t} - e^{-\lambda_{235}\cdot t}\right)\left(\frac{^{144}Nd}{^{238}U}\right)_{smp,P} + \left(\frac{^{235}U}{^{238}U}\right)_{smp,0}\frac{e^{-\lambda_{235}\cdot t}}{e^{-\lambda_{238}\cdot t}} \qquad (S.4)$$

Assuming that Cm is not chemically fractionated from Nd and that all samples form with the same initial U isotopic composition, Eq (S.4) can be rewritten,

$$\left(\frac{^{235}U}{^{238}U}\right)_{smp,P} = \left(\frac{^{247}Cm}{^{144}Nd}\right)_{solar,0} \left(\frac{\lambda_{247}}{\lambda_{235}-\lambda_{247}}\right)\left(e^{-\lambda_{247}\cdot t} - e^{-\lambda_{235}\cdot t}\right)\left(\frac{^{144}Nd}{^{238}U}\right)_{smp,P} + \left(\frac{^{235}U}{^{238}U}\right)_{solar,0} \frac{e^{-\lambda_{235}\cdot t}}{e^{-\lambda_{238}\cdot t}} \quad (S.5)$$

Introducing an arbitrary reference nuclide X ($^{235}$U in Fig. 1), one gets,

$$\left(\frac{^{235}U}{^{238}U}\right)_{smp,P} = \left(\frac{^{247}Cm}{X}\right)_{solar,0} \left(\frac{X}{^{144}Nd}\right)_{solar,0} \left(\frac{\lambda_{247}}{\lambda_{235}-\lambda_{247}}\right)\left(e^{-\lambda_{247}\cdot t} - e^{-\lambda_{235}\cdot t}\right)\left(\frac{^{144}Nd}{^{238}U}\right)_{smp,P} + \left(\frac{^{235}U}{^{238}U}\right)_{solar,0} \frac{e^{-\lambda_{235}\cdot t}}{e^{-\lambda_{238}\cdot t}} \quad (S.6)$$

Given that $^{247}$Cm has a short half-life compared to $^{235}$U, Eq. (S.6) can be simplified to

$$\left(\frac{^{235}U}{^{238}U}\right)_{smp,P} = \left(\frac{^{247}Cm}{X}\right)_{solar,0} \left(\frac{X}{^{144}Nd}\right)_{solar,0} e^{-\lambda_{235}\cdot t} \left(\frac{^{144}Nd}{^{238}U}\right)_{smp,P} + \left(\frac{^{235}U}{^{238}U}\right)_{solar,0} \frac{e^{-\lambda_{235}\cdot t}}{e^{-\lambda_{238}\cdot t}} \quad (S.7)$$

This last formula was used by earlier studies (*11, 33, 34*), however, it leads to a systematic error of 2.2 % on the initial $\left(\frac{^{247}Cm}{X}\right)_{solar,0}$, so we have decided to retain eq. (S6) in our calculations.

Equation S.6 describes the isochron that ESS samples should follow, had $^{247}$Cm been alive at the start of the SS. From the slope of the isochron (eq. S.8)

$$Slope = \left(\frac{^{247}Cm}{X}\right)_{solar,0} \left(\frac{X}{^{144}Nd}\right)_{solar,0} \left(\frac{\lambda_{247}}{\lambda_{235}-\lambda_{247}}\right)\left(e^{-\lambda_{247}\cdot t} - e^{-\lambda_{235}\cdot t}\right) \quad (S.8)$$

we can calculate the initial abundance of $^{247}$Cm at the time of closure of the samples, while the intercept (eq. S.9)

$$Intercept = \left(\frac{^{235}U}{^{238}U}\right)_{solar,0} \frac{e^{-\lambda_{235}\cdot t}}{e^{-\lambda_{238}\cdot t}} \quad (S.9)$$

provides the initial $^{235}$U/$^{238}$U ratio in the ESS.

Using our data, we obtain an initial ratio at the time of U depletion in the *Curious Marie* CAI of ($^{247}$Cm/$^{235}$U) = (5.6 ± 0.3) × 10$^{-5}$, which translates into a ($^{247}$Cm/$^{235}$U) ratio at solar system formation of (7.0 ± 1.6) × 10$^{-5}$ [equivalent to ($^{247}$Cm/$^{238}$U)$_{ESS}$ = (2.2 ± 0.5) × 10$^{-5}$ and ($^{247}$Cm/$^{232}$Th)$_{ESS}$ = (9.7 ± 2.2) ×

$10^{-6}$]. This initial estimate is consistent with the upper limit of $8 \times 10^{-5}$ derived from meteorite leachates (*33*, *34*).

**Galactic chemical evolution (GCE) model:**

*Production ratios*

GCE models rely on production ratios to predict the steady-state abundance of the elements in the ISM. The estimation of production ratios is difficult, especially for elements that are produced by the *r*-process as their predicted abundances depend on the nuclear mass function and anchoring of the model results to solar-system stable isotope abundances, which are missing for actinides (the closest stable isotope to the actinides is $^{209}$Bi).

To assess the robustness of the production ratios to be used in this study, we compiled and compared data from several recent studies (table S6). There is typically a factor of 2-3 difference between the various models. When the production ratios from the various studies agreed within an order of magnitude, we used the mid-range value as the production ratio value (*i.e.*, [Max + Min]/2) with the error reaching to the min and max values. Otherwise, a best estimate was taken, and an arbitrary uncertainty of 50% was assigned to this production ratio (see footnote of table S6 for details). In the case of $^{182}$Hf, we followed Meyer and Clayton (*6*): the best estimate production ratio ($^{182}$Hf/$^{177}$Hf = 0.55) was calculated using the decomposition into *s*- and *r*-process components of its daughter isotope $^{182}$W and the model of Bisterzo et al. 2014 (*3*).

*GCE modeling*

We use the open non-linear GCE model of Clayton (1988) (*29*) and Dauphas et al. (2003) (*28*), with k =1.7 and a presolar age of the galaxy of 8.7 Gyr. These parameters were chosen in order to reproduce the G-dwarf metallicity distribution. For all short-lived radionuclides (SLR), the production ratios used are the best estimates from table S6. Production ratios and initial solar system abundances are summarized in table S8. This data is graphically presented in Fig. 3.

Two astrophysical models can be used to interpret the data. In the simplest approach, the difference between the ratio $N_{SLR}/N_{Stable}$ in the interstellar medium (ISM) and early solar system is due to a time interval Δ, named "free-decay" interval, before solar system formation, during which the protosolar molecular cloud is isolated from further nucleosynthetic input and the various SLRs freely decay.

A second model, probably more physically meaningful, is the three-phase mixing model of Clayton 1983 (*30*). In this model, the product of supernovae explosions is injected into the warm ISM, which exchanges matter with large H1 clouds on a timescale $\tau_{12}$, which in turn exchanges matter with the protosolar molecular cloud on a given timescale $\tau_{23}$. In this model, the abundance of a SLR in the ESS is not controlled by the time since the last nucleosynthetic event, but by the exchange timescale between ISM and the molecular clouds. As the mixing timescale increases, the abundance of the SLR in the molecular cloud decreases, but not as rapidly as in the free decay interval model (for a given value of Δ = $\tau_{12} = \tau_{23}$).

Neither model can, however, explain the abundance of all SLR using a single free-decay interval or mixing timescale, as is evident from Fig. 3: $^{107}$Pd and $^{129}$I have similar $N_{SLR}/N_{Stable}$ ratios but different half-lives, making it impossible for both isotopes to be produced in a single event/process. In the "free-decay" interval framework, the simplest scenario explaining the abundance of all SLR requires the last injection of *p*-process, *s*-process and *r*-process nuclides to have occurred about 10 Myr, 35 Myr, and 100 Myr before solar system formation, respectively.

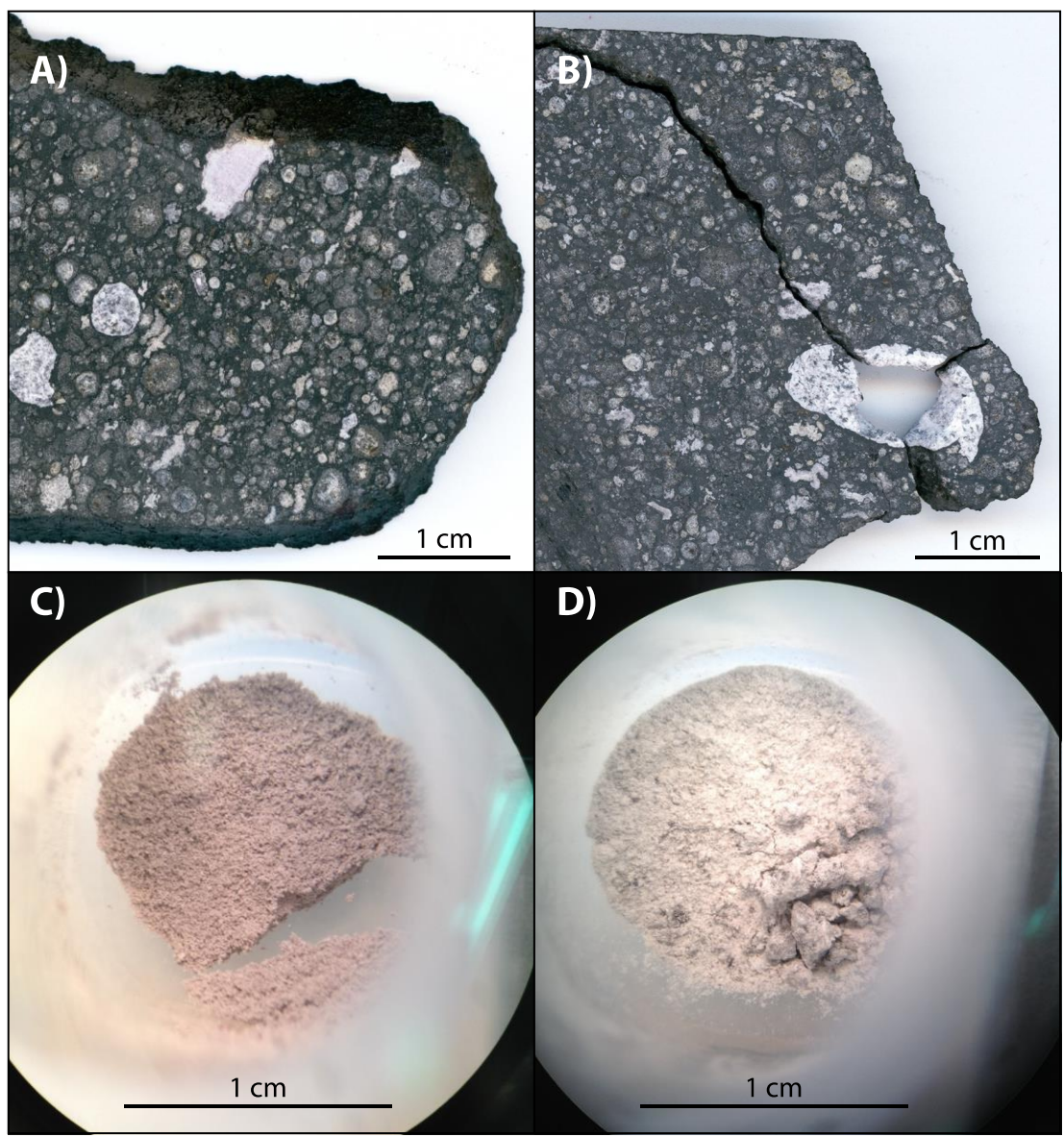

**fig. S1.** Pictures of typical fine-grained (left, FG-11) and coarse-grained (right, CG-2) CAIs. The top panels (a and b) show the CAIs still within the meteorite (Allende), and the bottom panels (c and d) show the powders in Teflon beakers after extraction (using cleaned stainless steel dental tools). See text for details about the texture, and mineralogy of fine- and coarse-grained CAIs.

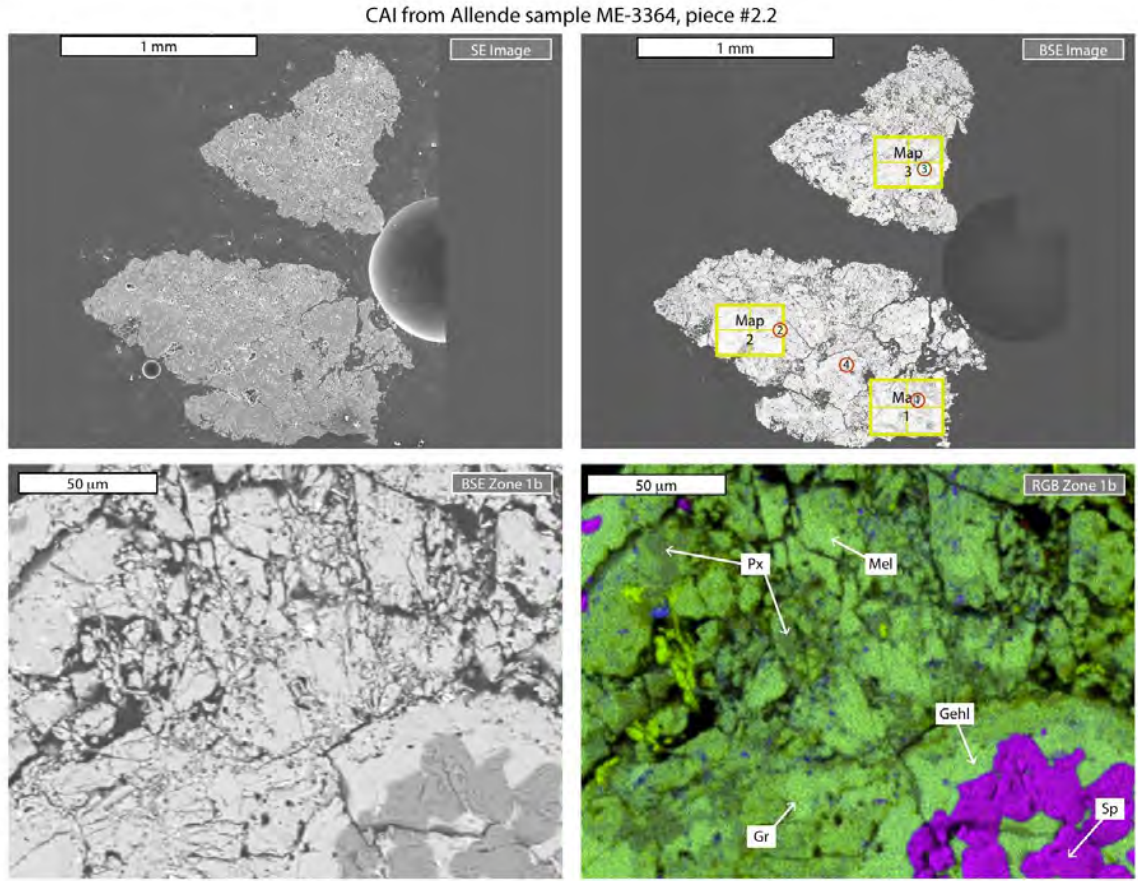

**fig. S2.**

Fig. S2 to S15: For each CAI studied in this work, Secondary Electron (SE) and Back Scattered Electron (BSE) images along with false color RGB (Mg/Ca/Al) maps of a selected field of view. The red numbered circles represent the location of laser ablation spots. The yellow areas (named Map X) show the zones mapped at higher magnification. Each yellow area is composed of 1 to 4 subzones, labeled by a letter (a to d) starting on the top left corner and moving clockwise. Note the different scales on the various maps. See text for more details about the texture, and mineralogy of fine- and coarse-grained CAIs.

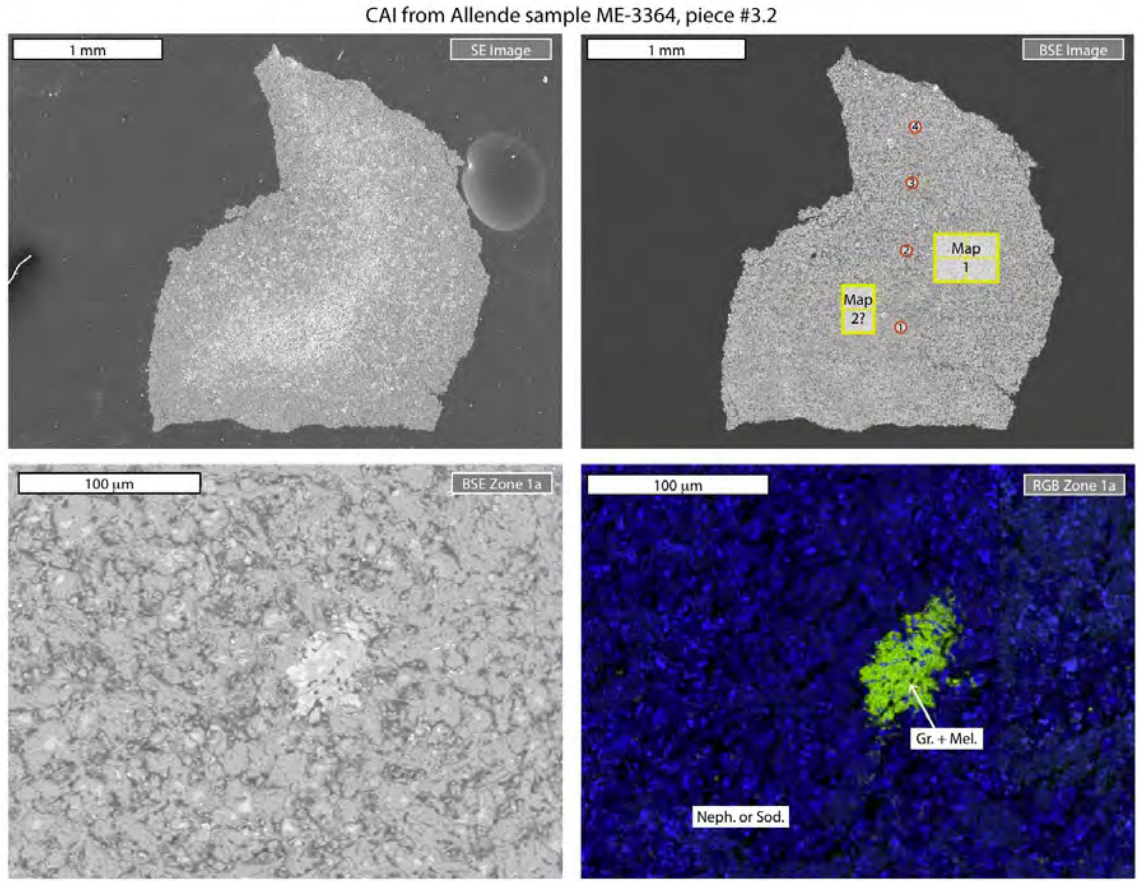

**fig. S3.**

Fig. S2 to S15: For each CAI studied in this work, Secondary Electron (SE) and Back Scattered Electron (BSE) images along with false color RGB (Mg/Ca/Al) maps of a selected field of view. The red numbered circles represent the location of laser ablation spots. The yellow areas (named Map X) show the zones mapped at higher magnification. Each yellow area is composed of 1 to 4 subzones, labeled by a letter (a to d) starting on the top left corner and moving clockwise. Note the different scales on the various maps. See text for more details about the texture, and mineralogy of fine- and coarse-grained CAIs.

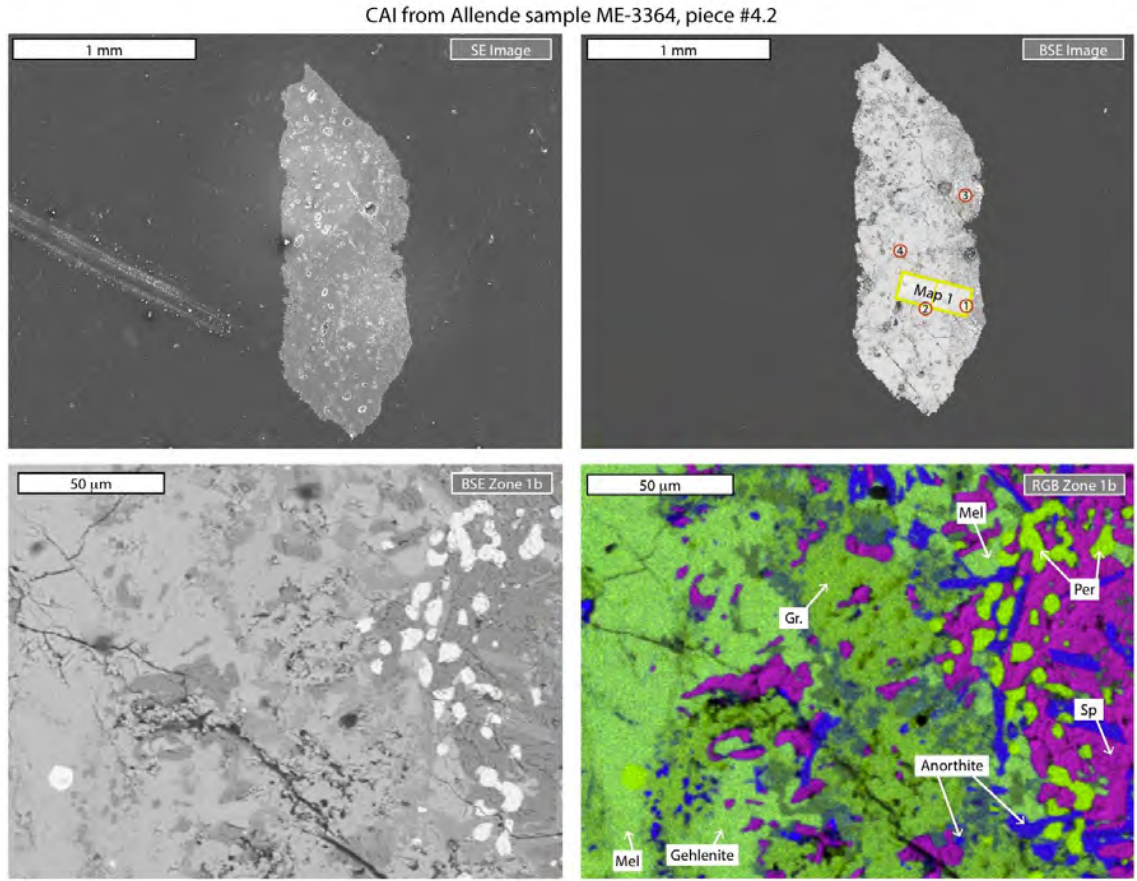

**fig. S4.**

Fig. S2 to S15: For each CAI studied in this work, Secondary Electron (SE) and Back Scattered Electron (BSE) images along with false color RGB (Mg/Ca/Al) maps of a selected field of view. The red numbered circles represent the location of laser ablation spots. The yellow areas (named Map X) show the zones mapped at higher magnification. Each yellow area is composed of 1 to 4 subzones, labeled by a letter (a to d) starting on the top left corner and moving clockwise. Note the different scales on the various maps. See text for more details about the texture, and mineralogy of fine- and coarse-grained CAIs.

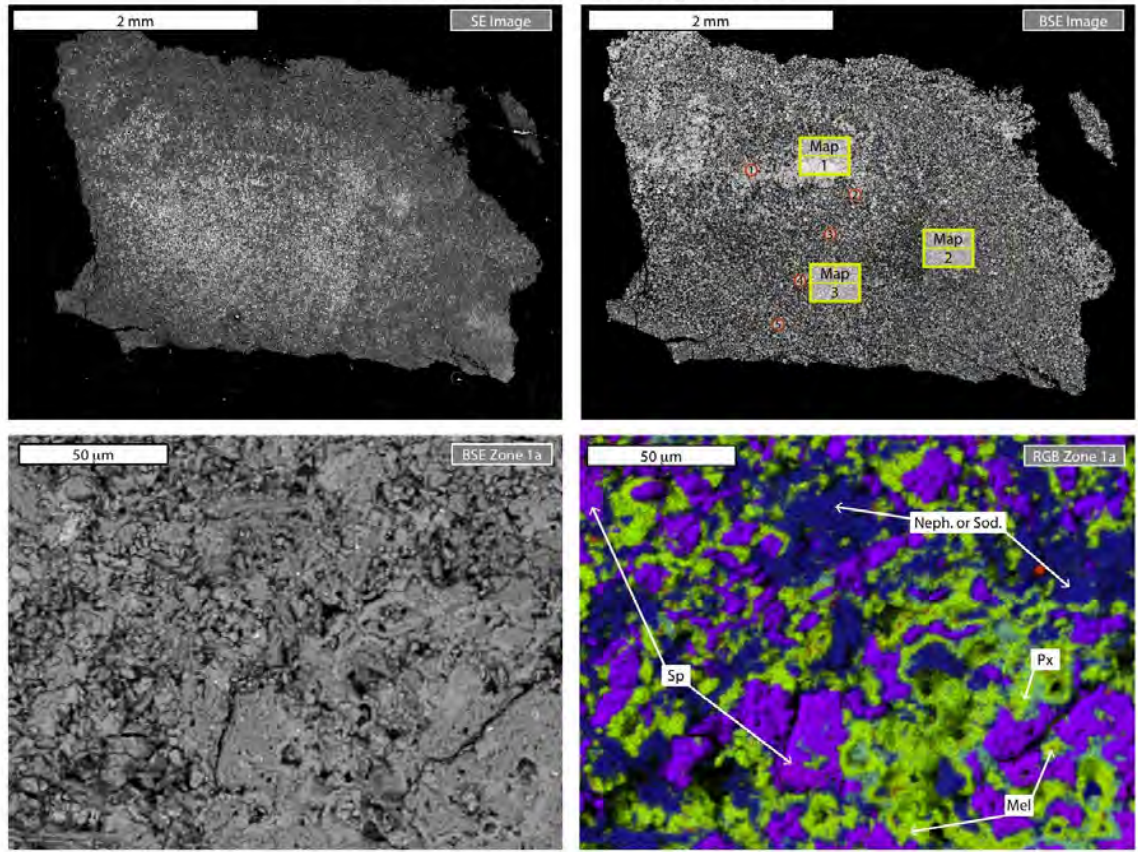

**fig. S5.**

Fig. S2 to S15: For each CAI studied in this work, Secondary Electron (SE) and Back Scattered Electron (BSE) images along with false color RGB (Mg/Ca/Al) maps of a selected field of view. The red numbered circles represent the location of laser ablation spots. The yellow areas (named Map X) show the zones mapped at higher magnification. Each yellow area is composed of 1 to 4 subzones, labeled by a letter (a to d) starting on the top left corner and moving clockwise. Note the different scales on the various maps. See text for more details about the texture, and mineralogy of fine- and coarse-grained CAIs.

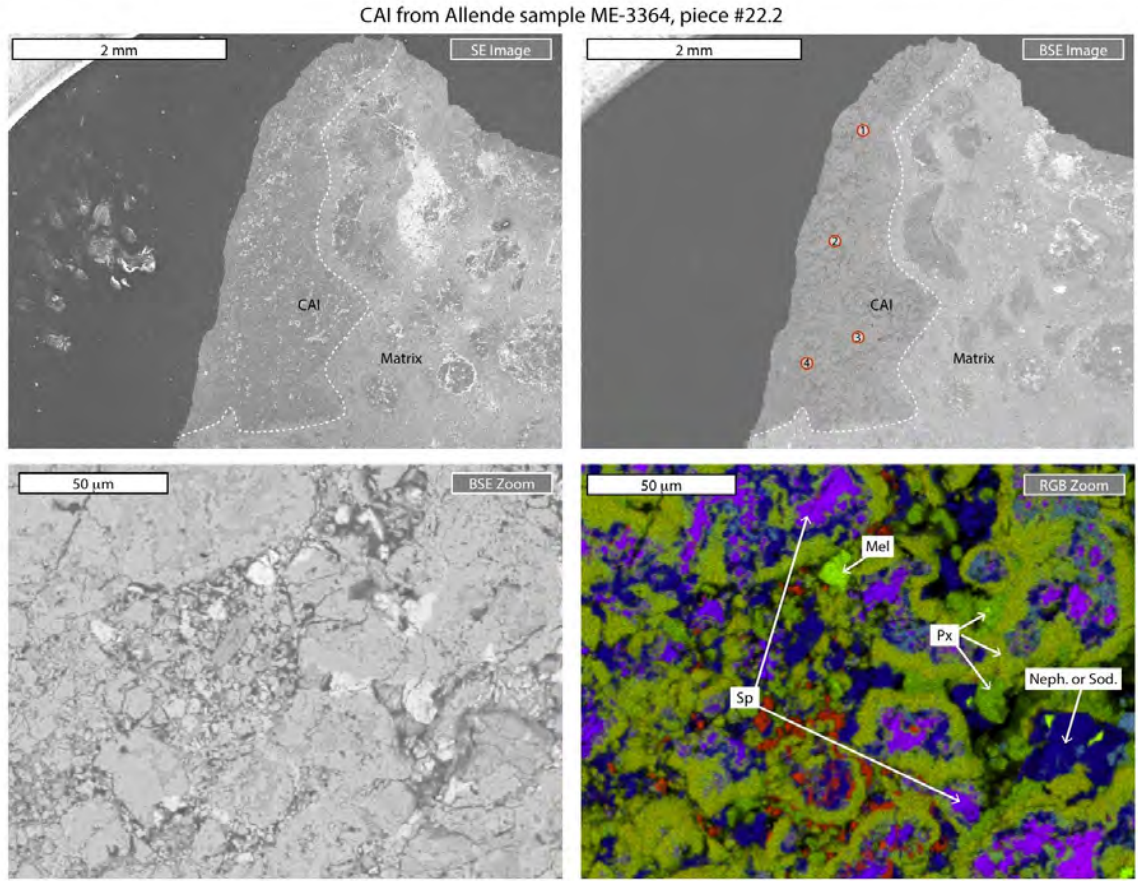

**fig. S6.**

Fig. S2 to S15: For each CAI studied in this work, Secondary Electron (SE) and Back Scattered Electron (BSE) images along with false color RGB (Mg/Ca/Al) maps of a selected field of view. The red numbered circles represent the location of laser ablation spots. The yellow areas (named Map X) show the zones mapped at higher magnification. Each yellow area is composed of 1 to 4 subzones, labeled by a letter (a to d) starting on the top left corner and moving clockwise. Note the different scales on the various maps. See text for more details about the texture, and mineralogy of fine- and coarse-grained CAIs.

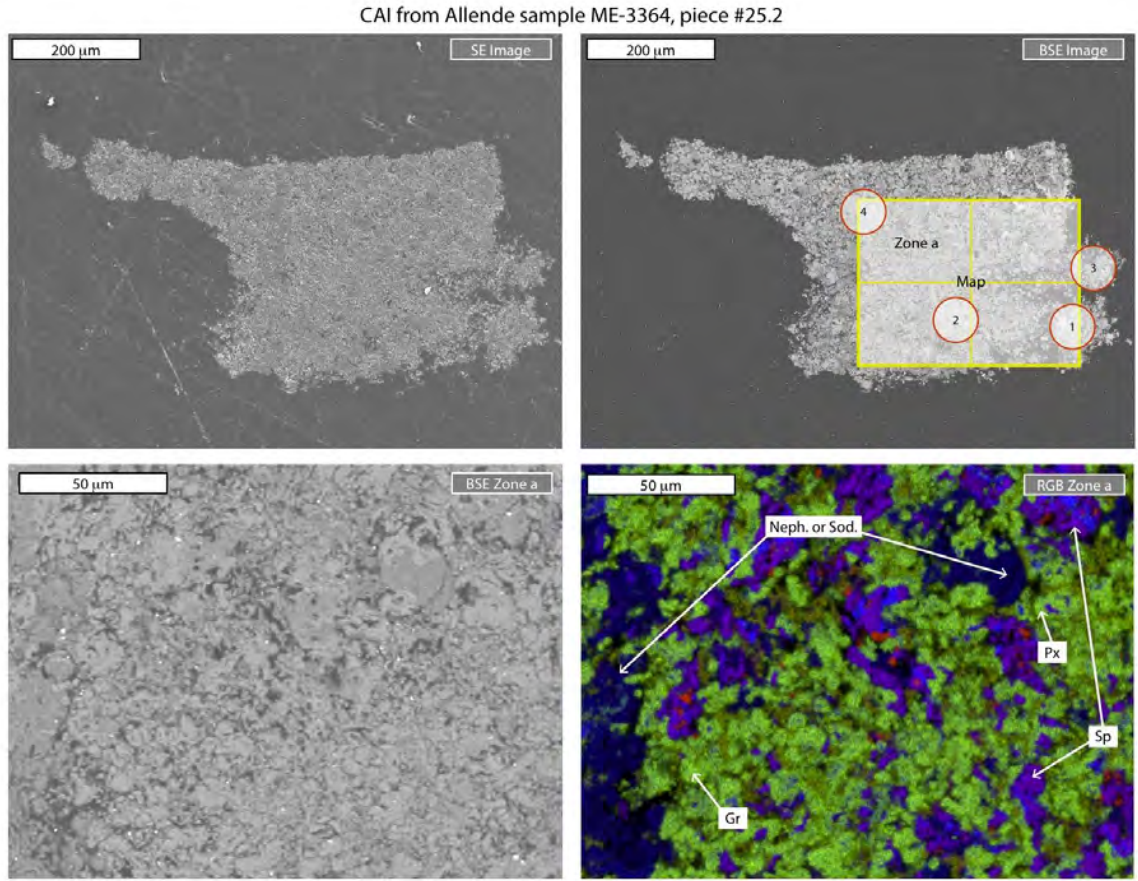

**fig. S7.**

Fig. S2 to S15: For each CAI studied in this work, Secondary Electron (SE) and Back Scattered Electron (BSE) images along with false color RGB (Mg/Ca/Al) maps of a selected field of view. The red numbered circles represent the location of laser ablation spots. The yellow areas (named Map X) show the zones mapped at higher magnification. Each yellow area is composed of 1 to 4 subzones, labeled by a letter (a to d) starting on the top left corner and moving clockwise. Note the different scales on the various maps. See text for more details about the texture, and mineralogy of fine- and coarse-grained CAIs.

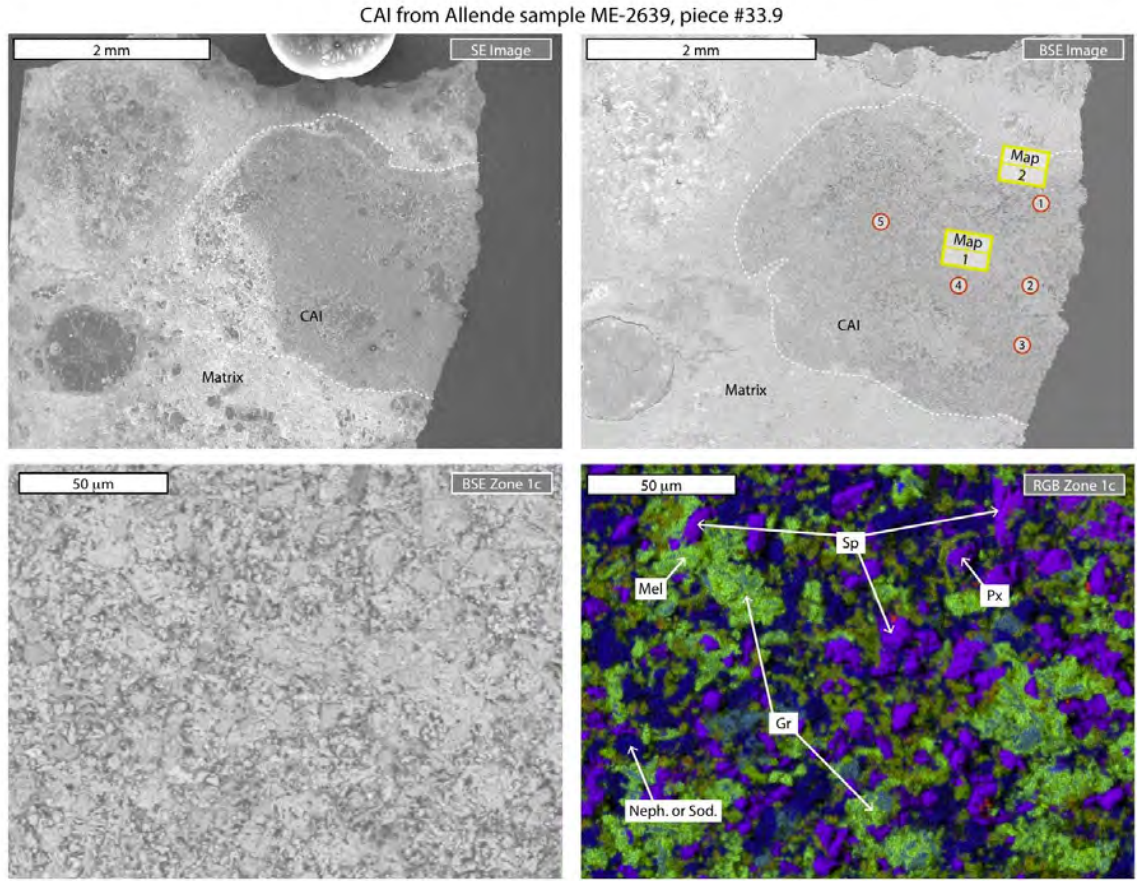

**fig. S8.**

Fig. S2 to S15: For each CAI studied in this work, Secondary Electron (SE) and Back Scattered Electron (BSE) images along with false color RGB (Mg/Ca/Al) maps of a selected field of view. The red numbered circles represent the location of laser ablation spots. The yellow areas (named Map X) show the zones mapped at higher magnification. Each yellow area is composed of 1 to 4 subzones, labeled by a letter (a to d) starting on the top left corner and moving clockwise. Note the different scales on the various maps. See text for more details about the texture, and mineralogy of fine- and coarse-grained CAIs.

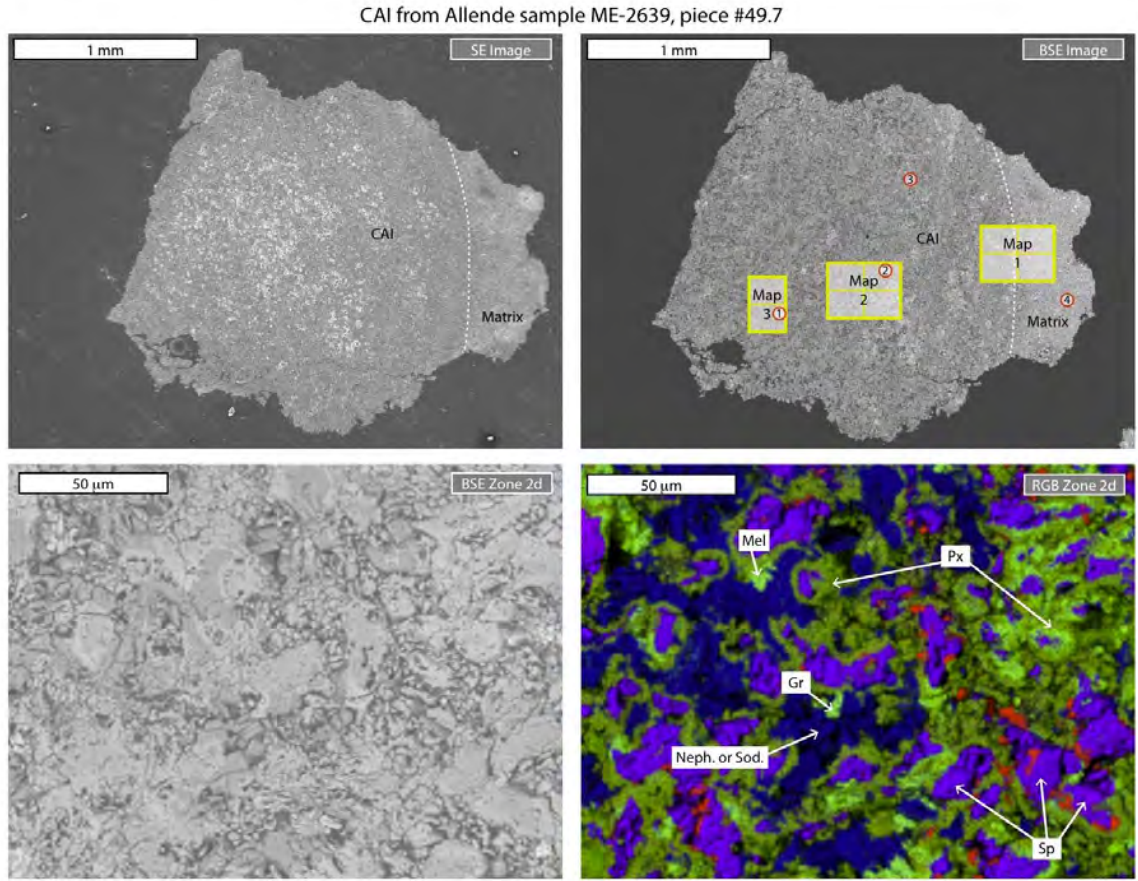

**fig. S9.**

Fig S2. to S15: For each CAI studied in this work, Secondary Electron (SE) and Back Scattered Electron (BSE) images along with false color RGB (Mg/Ca/Al) maps of a selected field of view. The red numbered circles represent the location of laser ablation spots. The yellow areas (named Map X) show the zones mapped at higher magnification. Each yellow area is composed of 1 to 4 subzones, labeled by a letter (a to d) starting on the top left corner and moving clockwise. Note the different scales on the various maps. See text for more details about the texture, and mineralogy of fine- and coarse-grained CAIs.

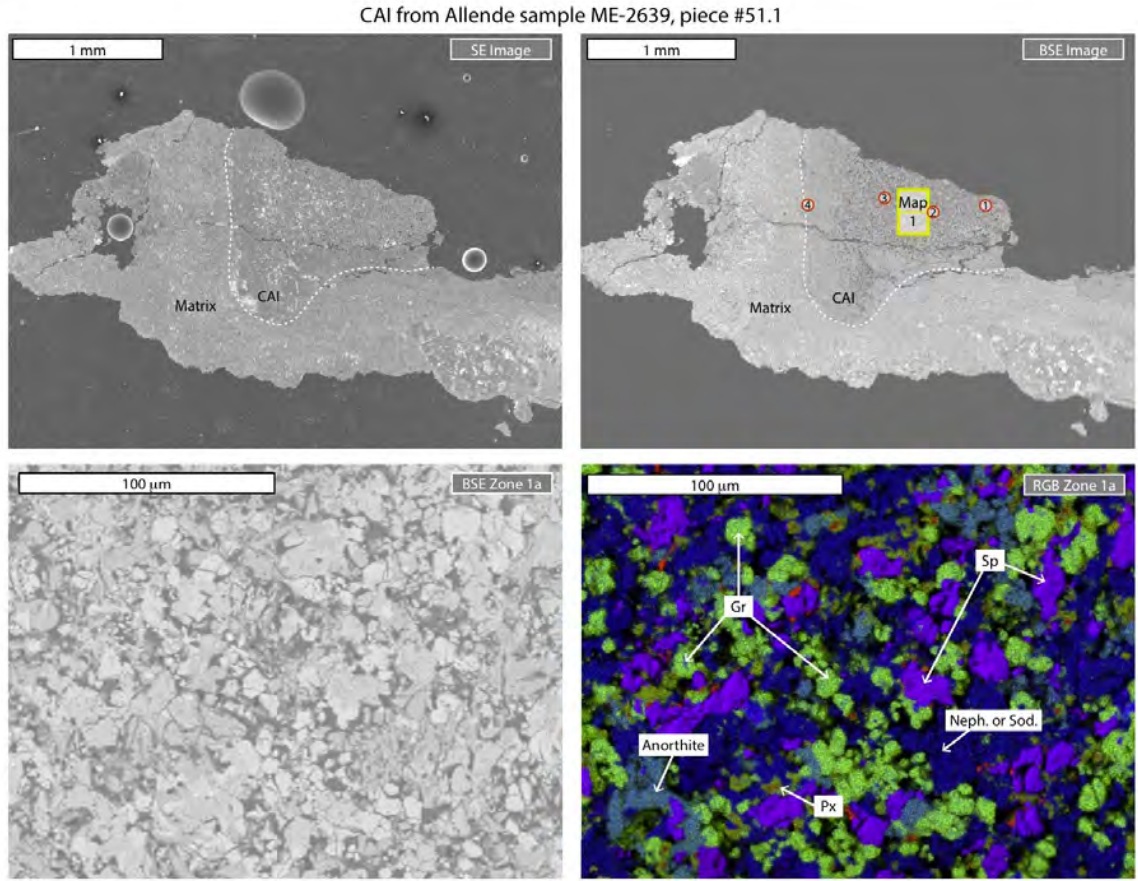

**fig. S10.**

Fig. S2 to S15: For each CAI studied in this work, Secondary Electron (SE) and Back Scattered Electron (BSE) images along with false color RGB (Mg/Ca/Al) maps of a selected field of view. The red numbered circles represent the location of laser ablation spots. The yellow areas (named Map X) show the zones mapped at higher magnification. Each yellow area is composed of 1 to 4 subzones, labeled by a letter (a to d) starting on the top left corner and moving clockwise. Note the different scales on the various maps. See text for more details about the texture, and mineralogy of fine- and coarse-grained CAIs.

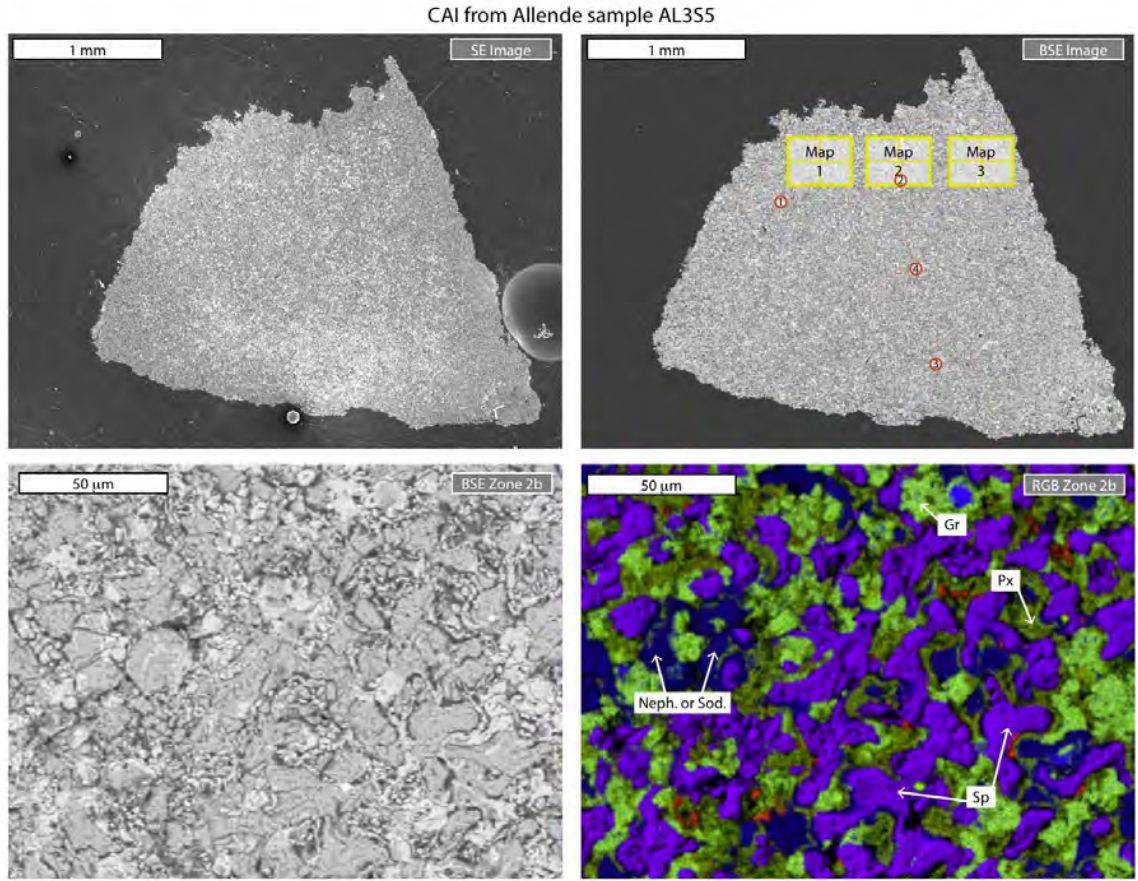

**fig. S11.**

Fig. S2 to S15: For each CAI studied in this work, Secondary Electron (SE) and Back Scattered Electron (BSE) images along with false color RGB (Mg/Ca/Al) maps of a selected field of view. The red numbered circles represent the location of laser ablation spots. The yellow areas (named Map X) show the zones mapped at higher magnification. Each yellow area is composed of 1 to 4 subzones, labeled by a letter (a to d) starting on the top left corner and moving clockwise. Note the different scales on the various maps. See text for more details about the texture, and mineralogy of fine- and coarse-grained CAIs.

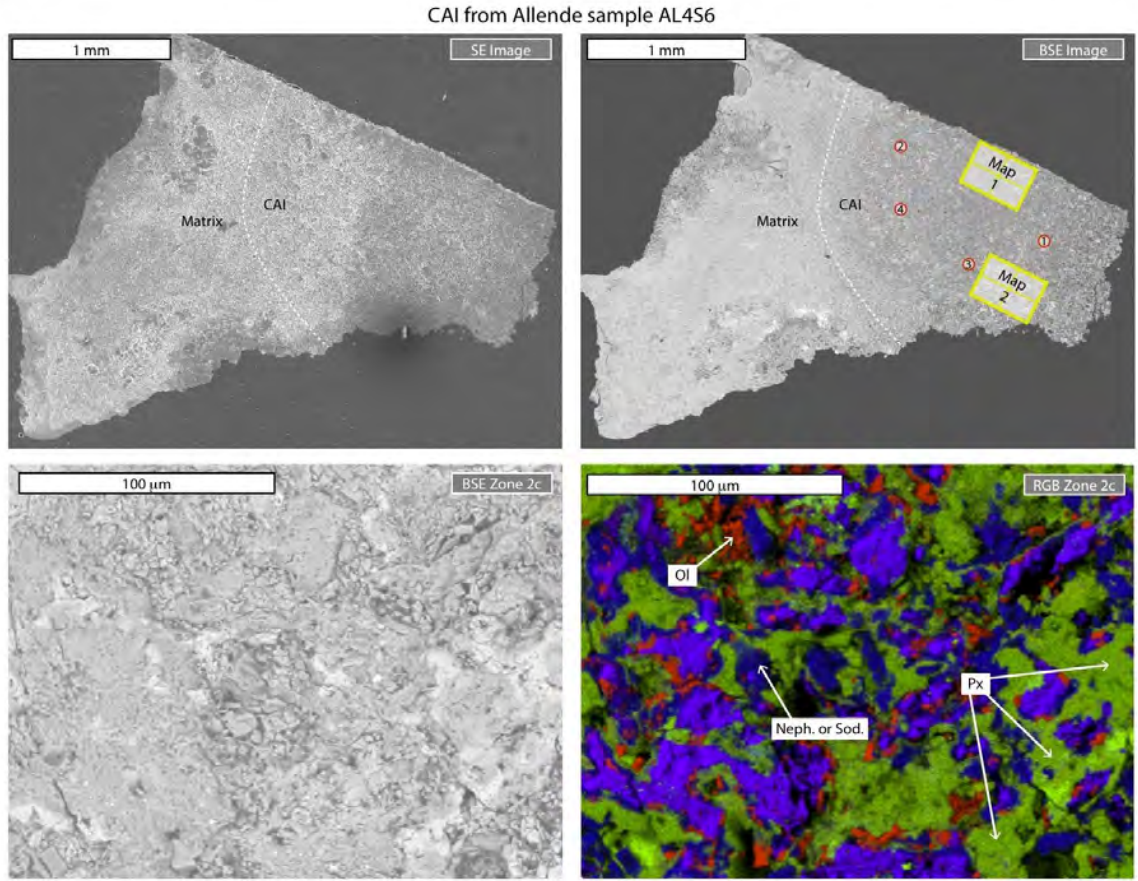

**fig. S12.**

Fig. S2 to S15: For each CAI studied in this work, Secondary Electron (SE) and Back Scattered Electron (BSE) images along with false color RGB (Mg/Ca/Al) maps of a selected field of view. The red numbered circles represent the location of laser ablation spots. The yellow areas (named Map X) show the zones mapped at higher magnification. Each yellow area is composed of 1 to 4 subzones, labeled by a letter (a to d) starting on the top left corner and moving clockwise. Note the different scales on the various maps. See text for more details about the texture, and mineralogy of fine- and coarse-grained CAIs.

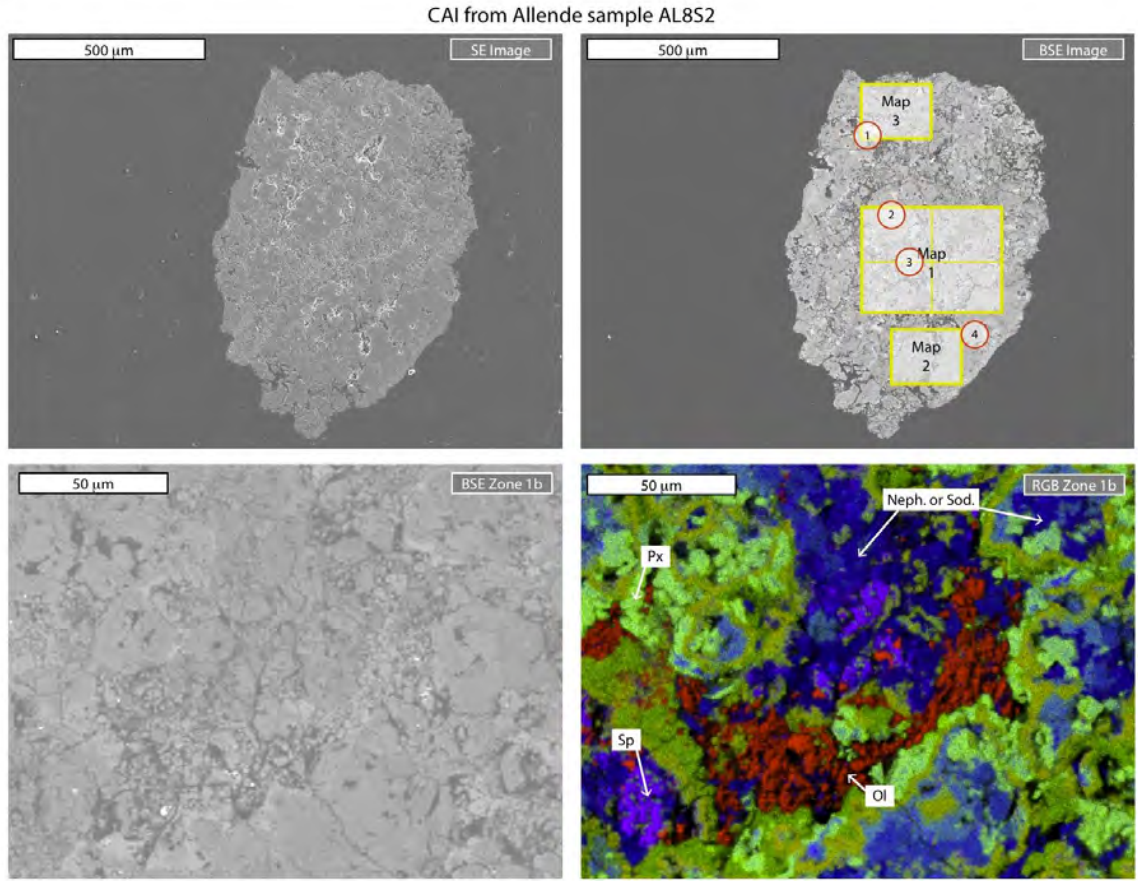

**fig. S13.**

Fig. S2 to S15: For each CAI studied in this work, Secondary Electron (SE) and Back Scattered Electron (BSE) images along with false color RGB (Mg/Ca/Al) maps of a selected field of view. The red numbered circles represent the location of laser ablation spots. The yellow areas (named Map X) show the zones mapped at higher magnification. Each yellow area is composed of 1 to 4 subzones, labeled by a letter (a to d) starting on the top left corner and moving clockwise. Note the different scales on the various maps. See text for more details about the texture, and mineralogy of fine- and coarse-grained CAIs.

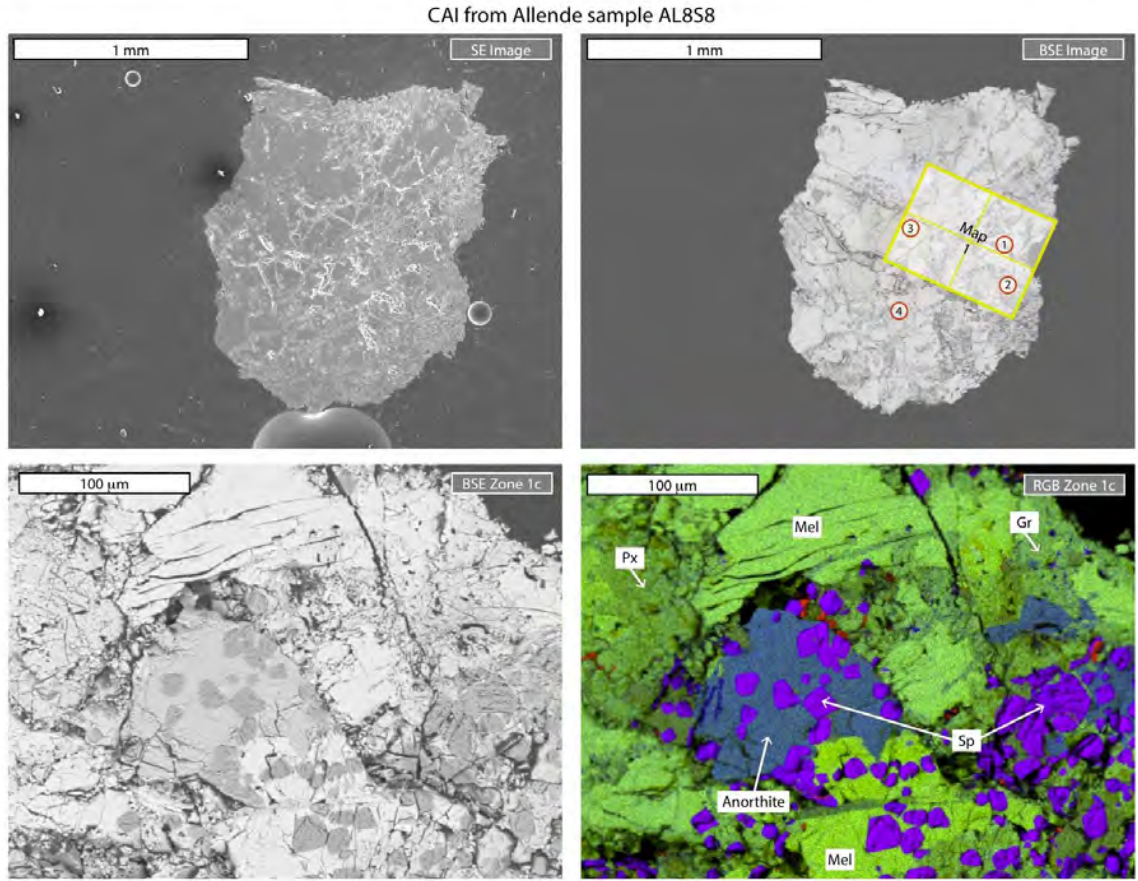

**fig. S14.**

Fig. S2 to S15: For each CAI studied in this work, Secondary Electron (SE) and Back Scattered Electron (BSE) images along with false color RGB (Mg/Ca/Al) maps of a selected field of view. The red numbered circles represent the location of laser ablation spots. The yellow areas (named Map X) show the zones mapped at higher magnification. Each yellow area is composed of 1 to 4 subzones, labeled by a letter (a to d) starting on the top left corner and moving clockwise. Note the different scales on the various maps. See text for more details about the texture, and mineralogy of fine- and coarse-grained CAIs.

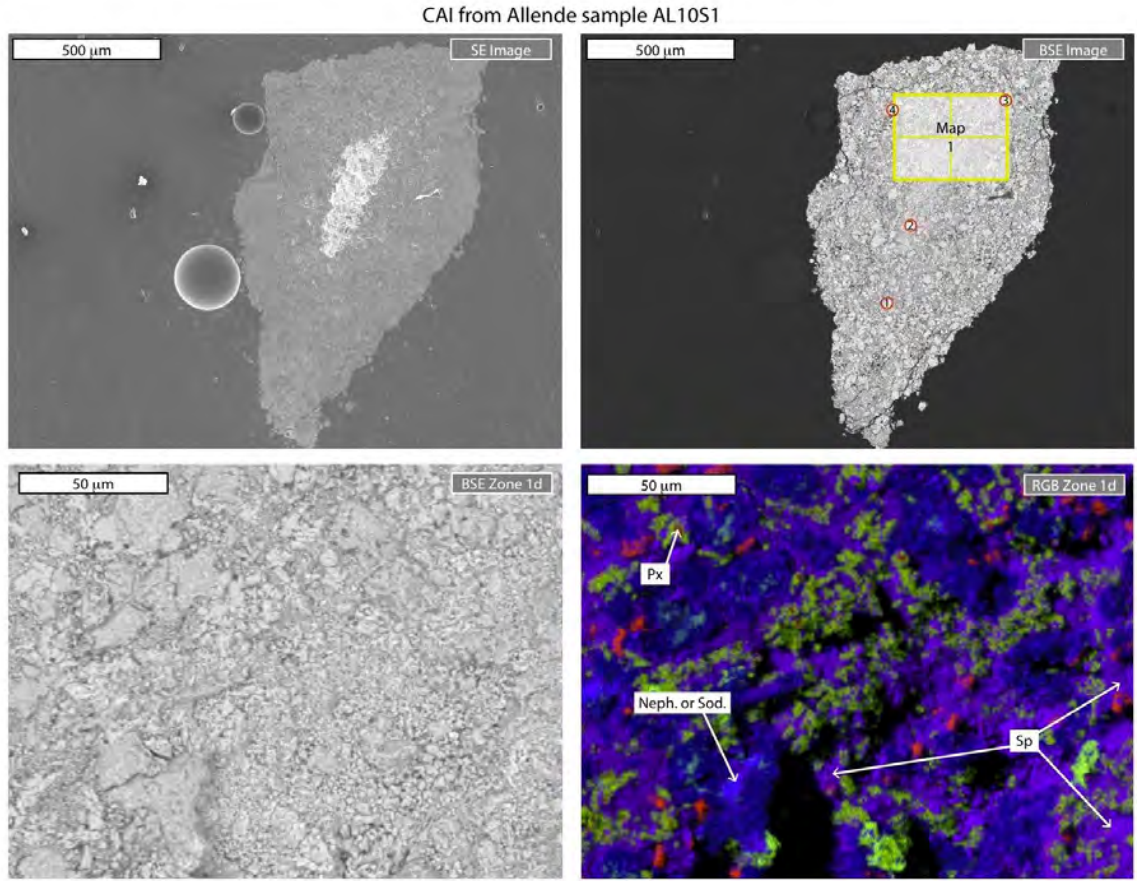

**fig. S15**

Fig. S2 to S15: For each CAI studied in this work, Secondary Electron (SE) and Back Scattered Electron (BSE) images along with false color RGB (Mg/Ca/Al) maps of a selected field of view. The red numbered circles represent the location of laser ablation spots. The yellow areas (named Map X) show the zones mapped at higher magnification. Each yellow area is composed of 1 to 4 subzones, labeled by a letter (a to d) starting on the top left corner and moving clockwise. Note the different scales on the various maps. See text for more details about the texture, and mineralogy of fine- and coarse-grained CAIs.

Table S1
Results of SEM analysis on small chips of CAIs mounted in epoxy (in wt %, normalized to 100 %).

| Nickname | Collection name | Na2O | MgO | Al2O3 | SiO2 | CaO | TiO2 | Fe2O3 |
|---|---|---|---|---|---|---|---|---|
| FG-1 | ME-3364-2.2 | 0.5 ± 0.7 | 2.9 ± 0.4 | 25.1 ± 3.1 | 34.6 ± 5.1 | 35.1 ± 2.7 | 0.4 ± 0.8 | 1.4 ± 1.1 |
| Curious Marie | ME-3364-3.2 | 14.8 ± 2.4 | 0.7 ± 0.2 | 44.0 ± 3.0 | 32.1 ± 1.6 | 3.4 ± 1.0 | 0.7 ± 0.3 | 4.2 ± 1.6 |
| CG-1 | ME-3364-4.2 | 0.6 ± 0.5 | 2.5 ± 2.8 | 34.8 ± 6.7 | 25.6 ± 5.4 | 34.9 ± 6.8 | 1.2 ± 2.2 | 0.4 ± 0.4 |
| FG-2 | ME-3364-22.2 | 3.5 ± 1.0 | 11.8 ± 3.8 | 19.8 ± 9.5 | 42.6 ± 8.0 | 17.5 ± 4.2 | 0.5 ± 0.8 | 4.3 ± 0.5 |
| FG-3 | ME-3364-25.2 | 4.0 ± 1.2 | 5.8 ± 3.0 | 26.7 ± 3.3 | 34.8 ± 1.6 | 21.4 ± 1.3 | 0.7 ± 0.2 | 6.6 ± 4.1 |
| FG-4 | ME-2639-16.2 | 5.2 ± 3.7 | 13.2 ± 2.4 | 37.6 ± 1.4 | 29.1 ± 1.6 | 8.6 ± 3.2 | 1.5 ± 0.4 | 4.9 ± 2.0 |
| FG-5 | ME-2639-33.9 | 3.2 ± 0.9 | 11.6 ± 1.4 | 34.1 ± 1.2 | 29.9 ± 0.7 | 13.7 ± 1.9 | 1.3 ± 0.3 | 6.1 ± 0.6 |
| FG-6 | ME-2639-49.7 | 4.0 ± 0.3 | 11.5 ± 0.6 | 26.7 ± 1.2 | 34.8 ± 0.7 | 15.1 ± 1.1 | 0.7 ± 0.4 | 7.1 ± 1.9 |
| FG-7 | ME-2639-51.1 | 4.5 ± 2.3 | 5.6 ± 2.1 | 30.9 ± 2.2 | 34.8 ± 1.5 | 19.6 ± 3.5 | 0.6 ± 0.4 | 3.9 ± 1.8 |
| FG-8 | AL3S5 | 3.4 ± 0.4 | 12.9 ± 0.5 | 37.9 ± 0.7 | 27.5 ± 1.2 | 14.2 ± 0.9 | 0.5 ± 0.3 | 3.6 ± 0.5 |
| FG-9 | AL4S6 | 4.7 ± 0.3 | 9.1 ± 2.1 | 24.8 ± 3.5 | 37.8 ± 2.4 | 15.1 ± 0.7 | 1.1 ± 0.4 | 7.6 ± 0.3 |
| FG-10 | AL8S2 | 3.5 ± 1.2 | 6.9 ± 1.9 | 22.8 ± 3.0 | 39.9 ± 1.2 | 19.6 ± 2.3 | 0.5 ± 0.2 | 6.9 ± 2.3 |
| CG-2 | AL8S8 | 0.2 ± 0.3 | 7.1 ± 1.5 | 26.4 ± 5.1 | 30.6 ± 3.6 | 34.2 ± 2.5 | 1.1 ± 1.5 | 0.3 ± 0.4 |
| FG-11 | AL10S1 | 6.7 ± 1.8 | 14.4 ± 2.0 | 44.1 ± 1.8 | 22.3 ± 2.3 | 4.7 ± 0.9 | 0.4 ± 0.2 | 7.5 ± 1.4 |

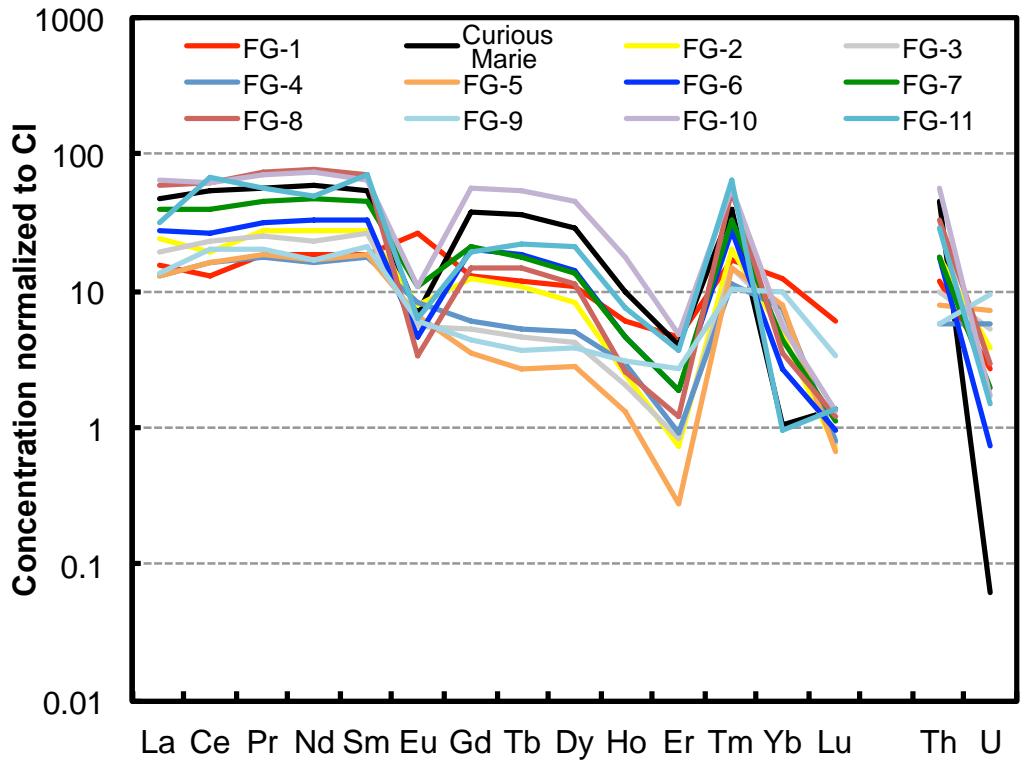

**fig. S16.** REE and U-Th abundance patterns of all twelve fine-grained CAIs analyzed in this study. All CAIs show a group II pattern, in which the most refractory (heavy REEs except Tm and Yb) and most volatile elements (Eu and Yb) are depleted relative to the moderately refractory light REEs. This pattern is thought to represent a snapshot in the condensation sequence as it corresponds to the composition of nebular dust condensates after depletion of the gas in ultra-refractory elements that condensed in hibonite (*57*) or perovskite (*23*), and before condensation of the more volatile ones that stayed behind in the gas (*22, 23*).

Table S2
Summary of U isotopic compositions and concentrations of CAIs and geostandards.

| Sample | Type | REE pattern | $^{144}$Nd/$^{238}$U | ± | Mass CAI used (mg) | Sample used (%) | ng U | Resin # | 235U Setup | Double Spike | n | Cycles | Data reduction | $\delta^{235}$U meas. (‰) | ± | $\delta[^{234}$U/$^{238}$U] (‰) | ± | Usp/Usmp | ± | $\delta^{235}$U blk[a] (‰) | ± | Blank contrib. | ± | [U] blk corr. (ppb) | ± | $\delta^{235}$U blk corr. (‰) | ± |
|---|---|---|---|---|---|---|---|---|---|---|---|---|---|---|---|---|---|---|---|---|---|---|---|---|---|---|---|
| **GEOSTANDARDS** | | | | | | | | | | | | | | | | | | | | | | | | | | | |
| BCR-2 | Basalt | | 7.1 | 0.7 | 11.39 | 100% | 18.34 | 1 | 10e12 Ω | ✓ | | 90 | MCS | 0.34 | 0.19 | 195.0 | 1.0 | 3.2% | | 0 | 10 | 0.16% | 0.04% | 1609 | 159 | 0.34 | 0.19 |
| | | | | | | | | 2 | 10e11 Ω | ✓ | | 42 | MCS | 0.15 | 0.28 | 0.6 | 0.8 | 3.2% | | 0 | 10 | 0.11% | 0.03% | 1611 | 154 | 0.15 | 0.28 |
| | | | | | | | | 3 | 10e11 Ω | ✓ | | 38 | MCS | 0.23 | 0.28 | 3.0 | 0.8 | 3.2% | | 0 | 10 | 0.12% | 0.03% | 1613 | 152 | 0.23 | 0.28 |
| | | | | | | | | | | | | | | | | | | | | | | | | **BCR-2 weighted average** | | **0.27** | **0.14** |
| BHVO-2 | Basalt | | 26.3 | 3.2 | 48.55 | 100% | 18.25 | 1 | 10e12 Ω | ✓ | | 90 | MCS | 0.67 | 0.19 | 51.9 | 1.0 | 3.1% | | 0 | 10 | 0.17% | 0.04% | 376 | 37 | 0.68 | 0.19 |
| | | | | | | | | 1 | 10e11 Ω | ✓ | | 80 | MCS | 0.24 | 0.30 | 45.1 | 0.4 | 3.1% | | 0 | 10 | 0.11% | 0.03% | 376 | 38 | 0.24 | 0.30 |
| | | | | | | | | | | | | | | | | | | | | | | | | **BHVO-2 weighted average** | | **0.55** | **0.16** |
| Faedra | Seawater | | | | 160710 | 100% | 0.767 | 1 | 10e12 Ω | ✓ | | 25 | MCS | 0.15 | 2.01 | 154.7 | 5.5 | 2.8% | | 0 | 10 | 0.08% | 0.02% | 21.5 | 2.6 | 0.15 | 2.01 |
| Allende | CV3 | | 20.2 | 2.5 | 1016.08 | 100% | 21.89 | 1 | 10e12 Ω | ✓ | | 90 | MCS | 0.61 | 0.19 | 71.3 | 1.0 | 2.1% | | 0 | 10 | 0.05% | 0.01% | 21.5 | 2.6 | 0.61 | 0.19 |
| | | | | | | | | 1 | 10e11 Ω | ✓ | | 88 | MCS | 0.21 | 0.30 | 46.7 | 0.4 | 2.1% | | 0 | 10 | | | | | 0.21 | 0.30 |
| | | | | | | | | | | | | | | | | | | | | | | | | **Allende weighted average** | | **0.49** | **0.16** |
| **CAIs** | | | | | | | | | | | | | | | | | | | | | | | | | | | |
| FG-1a | Fine-gr. CAI | Group II | 161.1 | 16.3 | 29 | 79.9% | 0.62 | 1 | 10e12 Ω | ✓ | | 26 | MCS | 1.93 | 1.55 | 66.7 | 5.5 | 7.2% | | 0 | 10 | 0.98% | 0.24% | 21.1 | 0.4 | 1.95 | 1.57 |
| FG-1b | | | 152.3 | 16.4 | 6.0 | 16.5% | 0.15 | 1 | 10e12 Ω | ✓ | | 23 | MCS | 5.27 | 7.64 | 1545.5 | 7.4 | 4.6% | | 0 | 10 | 8.21% | 0.63% | 22.3 | 0.9 | 5.74 | 8.38 |
| | **Weighted average** | | 156.7 | 11.6 | | | | | | | | | | | | | | | | | | | | **FG-1 weighted average** | | **2.08** | **1.54** |
| Curious Marie a | Fine-gr. CAI | Group II | 22647 | 805 | 594 | 77.4% | 0.29 | 1 | 10e12 Ω | ✓ | | 29 | MCS | 52.12 | 14.72 | 1213.1 | 18.4 | 0.61% | | 0 | 10 | 1.27% | 0.38% | 0.482 | 0.008 | 52.79 | 14.91 |
| Curious Marie b | | | 21446 | 3648 | 80 | 10.4% | 0.048 | 2 | 10e12 Ω | ✓ | | 41 | Excel | 51.59 | 1.26 | | | | | 0 | 10 | 12.73% | 2.82% | 0.51 | 0.09 | 59.12 | 2.80 |
| Curious Marie c | | | 25761 | 6289 | 44 | 5.8% | 0.032 | 3 | SEM | ✓ | | 31 | Excel | 42.09 | 1.46 | | | | | 1.61 | 3.58 | 29.43% | 2.31% | 0.42 | 0.10 | 58.97 | 3.17 |
| | **Weighted average** | | 22640 | 780 | | | | | | | | | | | | | | | | | | | | **Curious Marie weighted average** | | **58.93** | **2.08** |
| CG-1a | Coarse-gr. CAI | Group I | 122.5 | 12.4 | 40 | 79.7% | 0.61 | 1 | 10e12 Ω | ✓ | | 29 | MCS | 1.35 | 1.52 | 122.6 | 5.4 | 7.3% | | 0 | 10 | 0.94% | 0.24% | 15.2 | 0.2 | 1.36 | 1.54 |
| CG-1b | | | 119.1 | 12.4 | 8.4 | 16.7% | 0.14 | 1 | 10e12 Ω | ✓ | | 32 | MCS | 0.89 | 7.64 | 110.0 | 7.4 | 4.6% | | 0 | 10 | 7.50% | 1.09% | 15.6 | 0.5 | 0.96 | 8.30 |
| | **Weighted average** | | 120.8 | 8.8 | | | | | | | | | | | | | | | | | | | | **CG-1 weighted average** | | **1.35** | **1.52** |
| FG-2a | Fine-gr. CAI | Group II | 178.8 | 9.0 | 42 | 79.7% | 1.24 | 1 | 10e12 Ω | ✓ | | 18 | MCS | 0.23 | 0.73 | 27.4 | 2.6 | 7.2% | | 0 | 10 | 0.39% | 0.10% | 29.7 | 0.3 | 0.23 | 0.73 |
| FG-2b | | | 168.6 | 9.3 | 8.9 | 17.1% | 0.29 | 1 | 10e12 Ω | ✓ | | 21 | MCS | -0.75 | 2.01 | 76.6 | 5.5 | 4.8% | | 0 | 10 | 3.68% | 0.28% | 31.5 | 0.8 | -0.78 | 2.12 |
| | **Weighted average** | | 173.9 | 6.5 | | | | | | | | | | | | | | | | | | | | **FG-2 weighted average** | | **0.12** | **0.69** |
| FG-3a | Fine-gr. CAI | Group II | 108.4 | 6.5 | 81 | 79.8% | 3.29 | 1 | 10e12 Ω | ✓ | | 77 | MCS | 1.67 | 0.60 | 21.3 | 1.2 | 6.1% | | 0 | 10 | 0.33% | 0.08% | 40.5 | 0.2 | 1.68 | 0.60 |
| FG-3b | | | 101.6 | 6.3 | 17 | 17.1% | 0.76 | 1 | 10e12 Ω | ✓ | | 24 | MCS | -0.48 | 0.88 | 55.8 | 2.3 | 4.9% | | 0 | 10 | 1.52% | 0.12% | 43.2 | 0.6 | -0.49 | 0.90 |
| | **Weighted average** | | 104.8 | 4.5 | | | | | | | | | | | | | | | | | | | | **FG-3 weighted average** | | **1.01** | **0.50** |
| FG-4 | Fine-gr. CAI | Group II | 66.9 | 6.7 | 350 | 79.8% | 15.74 | 1 | 10e12 Ω | ✓ | | 84 | MCS | -0.89 | 0.35 | 10.4 | 1.3 | 5.3% | | 0 | 10 | 0.10% | 0.03% | 45.01 | 0.06 | -0.89 | 0.35 |
| | | | 66.8 | 6.7 | | | | 2 | 10e12 Ω | ✓ | | 90 | MCS | -0.77 | 0.35 | 25.2 | 1.3 | 5.3% | | 0 | 10 | | | | | -0.77 | 0.35 |
| FG-5a | Fine-gr. CAI | Group II | 55.9 | 5.6 | 40 | 79.7% | 2.29 | 1 | 10e12 Ω | ✓ | | 41 | MCS | 0.59 | 0.64 | 30.1 | 1.7 | 7.3% | | 0 | 10 | 0.32% | 0.08% | 56.9 | 0.5 | 0.60 | 0.64 |
| FG-5b | | | 52.6 | 5.4 | 8.6 | 17.2% | 0.53 | 1 | 10e12 Ω | ✓ | | 22 | MCS | -1.28 | 1.44 | 54.2 | 3.9 | 4.9% | | 0 | 10 | 1.98% | 0.15% | 60.5 | 1.6 | -1.30 | 1.49 |
| | **Weighted average** | | 54.2 | 3.9 | | | | | | | | | | | | | | | | | | | | **FG-5 weighted average** | | **0.30** | **0.59** |
| FG-6a | Fine-gr. CAI | Group II | 1060 | 73 | 45 | 79.7% | 0.27 | 1 | 10e12 Ω | ✓ | | 27 | MCS | 0.37 | 2.19 | -5.0 | 3.0 | 6.7% | | 0 | 10 | 1.78% | 0.44% | 5.8 | 0.06 | 0.38 | 2.23 |
| FG-6b | | | 828 | 138 | 9.7 | 17.1% | 0.085 | 3 | SEM | ✓ | | 31 | Excel | 3.58 | 1.03 | | | | | 1.61 | 3.58 | 14.02% | 1.02% | 7.4 | 1.1 | 3.90 | 1.33 |
| | **Weighted average** | | 1009 | 64 | | | | | | | | | | | | | | | | | | | | **FG-6 weighted average** | | **2.98** | **1.14** |
| FG-7a | Fine-gr. CAI | Group II | 567 | 22 | 51 | 79.7% | 0.79 | 1 | 10e12 Ω | ✓ | | 30 | MCS | 2.13 | 0.96 | 26.7 | 3.1 | 6.7% | | 0 | 10 | 0.66% | 0.17% | 15.5 | 0.2 | 2.15 | 0.96 |
| FG-7b | | | 540 | 23 | 11 | 16.6% | 0.18 | 1 | 10e12 Ω | ✓ | | 26 | MCS | 1.03 | 7.64 | 414.1 | 7.4 | 4.6% | | 0 | 10 | 7.74% | 0.61% | 16.3 | 0.4 | 1.11 | 8.35 |
| | **Weighted average** | | 555 | 16 | | | | | | | | | | | | | | | | | | | | **FG-7 weighted average** | | **2.13** | **0.96** |
| FG-8 | Fine-gr. CAI | Group II | 653 | 14 | 377 | 79.8% | 8.50 | 1 | 10e12 Ω | ✓ | | 90 | MCS | 3.39 | 0.35 | 277.3 | 1.3 | 4.4% | | 0 | 10 | 0.11% | 0.03% | 22.53 | 0.04 | 3.40 | 0.35 |
| | | | 652 | 14 | | | | 2 | 10e12 Ω | ✓ | | 22 | MCS | 3.28 | 0.26 | 102.7 | 2.0 | 4.4% | | 0 | 10 | | | | | 3.28 | 0.26 |
| FG-9 | Fine-gr. CAI | Group II | 653 | 10 | 331 | 79.8% | 23.93 | 1 | 10e12 Ω | ✓ | | 78 | MCS | 1.90 | 0.19 | 17.4 | 1.0 | 5.5% | | 0 | 10 | 0.07% | 0.02% | 72.35 | 0.10 | 1.90 | 0.19 |
| | | | | | | | | 2 | 10e12 Ω | ✓ | | 89 | MCS | 2.00 | 0.19 | 34.2 | 1.0 | 5.5% | | 0 | 10 | | | | | 2.00 | 0.19 |
| | | | | | | | | | | | | | | | | | | | | | | | | **FG-8 weighted average** | | **3.32** | **0.21** |
| | | | | | | | | | | | | | | | | | | | | | | | | **FG-9 weighted average** | | **1.95** | **0.14** |
| FG-10a | Fine-gr. CAI | Group II | 1057 | 64 | 13 | 79.7% | 0.18 | 1 | 10e12 Ω | ✓ | | 23 | MCS | 0.46 | 2.19 | 900.2 | 3.0 | 5.8% | | 0.00 | 10 | 2.84% | 0.71% | 13.3 | 0.4 | 0.48 | 2.27 |
| FG-10b | | | 834 | 156 | 2.6 | 16.1% | 0.058 | 3 | SEM | ✓ | | 29 | MCS | 2.84 | 1.21 | | | | | 1.61 | 3.58 | 19.75% | 1.69% | 16.9 | 3.0 | 3.14 | 1.75 |
| | | | | | | | | | | | | | | | | | | | | | | | | **FG-10 weighted average** | | **3.14** | **1.75** |
| CG-2 | Coarse-gr. CAI | Group V | 31.5 | 3.2 | 200 | 79.8% | 22.99 | 1 | 10e12 Ω | ✓ | | 86 | MCS | 0.43 | 0.19 | 20.5 | 1.0 | 6.5% | | 0 | 10 | 0.07% | 0.02% | 114.7 | 0.2 | 0.43 | 0.19 |
| | | | 31.5 | 3.2 | | | | 2 | 10e12 Ω | ✓ | | 90 | MCS | 0.43 | 0.19 | 13.2 | 1.0 | 6.5% | | 0 | 10 | | | | | 0.43 | 0.19 |
| | | | | | | | | | | | | | | | | | | | | | | | | **CG-2 weighted average** | | **0.43** | **0.14** |
| FG-11a | Fine-gr. CAI | Group II | 787 | 22 | 57 | 79.7% | 0.67 | 1 | 10e12 Ω | ✓ | | 20 | MCS | 5.98 | 1.02 | 62.8 | 5.0 | 6.8% | | 0 | 10 | 0.73% | 0.18% | 11.8 | 0.3 | 6.03 | 1.03 |
| FG-11b | | | 745 | 24 | 12 | 17.1% | 0.16 | 1 | 10e12 Ω | ✓ | | 19 | MCS | 6.06 | 7.64 | 210.5 | 7.4 | 4.7% | | 0 | 10 | 7.91% | 0.61% | 12.4 | 0.3 | 6.58 | 8.35 |
| | **Weighted average** | | 768 | 16 | | | | | | | | | | | | | | | | | | | | **FG-11 weighted average** | | **6.03** | **1.02** |
| TS32a | Coarse-gr. CAI | Group V | 34.0 | 3.4 | 34 | 79.6% | 3.96 | 1 | 10e12 Ω | ✓ | | 46 | MCS | 0.00 | 0.32 | -59.2 | 1.7 | 7.2% | | 0 | 10 | 0.28% | 0.07% | 115.0 | 1.1 | 0.00 | 0.32 |
| TS32b | | | 31.6 | 3.3 | 7.4 | 17.2% | 0.93 | 1 | 10e12 Ω | ✓ | | 21 | MCS | -1.41 | 0.88 | -39.1 | 2.3 | 4.8% | | 0 | 10 | 1.20% | 0.09% | 123.7 | 3.9 | -1.43 | 0.89 |
| | **Weighted average** | | 32.8 | 2.4 | | | | | | | | | | | | | | | | | | | | **TS32 weighted average** | | **-0.16** | **0.30** |
| **DOPING TESTS** | | | | | | | | | | | | | | | | | | | | | | | | | | | |
| Sample matrix | Doping standard | | | | | | | | | | | | | | | | | | | | | | | | | | |
| FG-1 | CRM-112a | | | | | | 0.49 | | 10e12 Ω | ✓ | | 23 | MCS | -0.74 | 1.18 | -22.3 | 2.5 | 6.2% | | 0 | 10 | 0.97% | 0.14% | | | -0.75 | 1.20 |
| Curious Marie | CRM-112a | | | | | | 0.34 | | 10e12 Ω | ✓ | | 32 | MCS | 2.10 | 16.77 | 65.3 | 22.7 | 0.51% | | 0 | 10 | | | | | 2.10 | 16.77 |
| CG-1 | CRM-112a | | | | | | 0.59 | | 10e12 Ω | ✓ | | 25 | MCS | 1.02 | 1.18 | 147.3 | 2.5 | 6.3% | | 0 | 10 | 0.94% | 0.13% | | | 1.03 | 1.20 |
| FG-2 | CRM-112a | | | | | | 1.17 | | 10e12 Ω | ✓ | | 25 | MCS | -0.18 | 1.18 | 74.5 | 2.5 | 6.3% | | 0 | 10 | 0.70% | 0.10% | | | -0.18 | 1.19 |
| FG-3 | CRM-112a | | | | | | 3.14 | | 10e12 Ω | ✓ | | 41 | MCS | 0.03 | 0.53 | 153.4 | 2.8 | 6.3% | | 0 | 10 | 0.45% | 0.06% | | | 0.03 | 0.54 |
| FG-5 | CRM-112a | | | | | | 2.21 | | 10e12 Ω | ✓ | | 24 | MCS | -0.06 | 0.53 | 4.0 | 2.8 | 6.2% | | 0 | 10 | 0.44% | 0.06% | | | -0.06 | 0.54 |
| FG-6 | CRM-112a | | | | | | 0.26 | | 10e12 Ω | ✓ | | 30 | MCS | -0.33 | 2.72 | 1.1 | 3.9 | 6.2% | | 0 | 10 | | | | | -0.33 | 2.72 |
| FG-7 | CRM-112a | | | | | | 0.84 | | 10e12 Ω | ✓ | | 24 | MCS | 0.38 | 1.18 | -37.2 | 2.5 | 6.3% | | 0 | 10 | 0.71% | 0.10% | | | 0.38 | 1.19 |
| FG-10a | CRM-112a | | | | | | 0.18 | | 10e12 Ω | ✓ | | 26 | MCS | -41.20 | 2.72 | -63.9 | 3.9 | 5.9% | | 0 | 10 | | | | | -41.20 | 2.72 |
| FG-10b | CRM-112a | | | | | | 0.18 | | 10e12 Ω | ✓ | | 32 | MCS | -17.89 | 3.19 | -39.5 | 5.2 | 5.5% | | 0 | 10 | 2.29% | 0.32% | | | -18.31 | 3.28 |
| FG-11a | CRM-112a | | | | | | 0.69 | | 10e12 Ω | ✓ | | 32 | MCS | 1.59 | 2.72 | 90.8 | 3.9 | 6.3% | | 0 | 10 | | | | | 1.59 | 2.72 |
| FG-11b | CRM-112a | | | | | | 0.69 | | 10e12 Ω | ✓ | | 24 | MCS | 0.03 | 1.18 | -34.7 | 2.5 | 6.3% | | 0 | 10 | 0.84% | 0.12% | | | 0.03 | 1.19 |
| TS32 | CRM-112a | | | | | | 3.77 | | 10e12 Ω | ✓ | | 34 | MCS | 0.46 | 0.53 | 92.3 | 2.3 | 6.3% | | 0 | 10 | 0.36% | 0.06% | | | 0.46 | 0.53 |

[a] In most cases, the 235U signal was too low to determine the d235U of the chemistry blank reliably. An arbitrary value of 0 ± 10 ‰ was used. This approach is conservative as the 10 ‰ uncertainty is propagated into the final estimate of the blank corrected d235U value.
[b] The double spike data reduction was done using a Mathematica code [Tissot & Dauphas (2015)] and error propagation by MonteCarlo Simulations (MCS). Data of unspiked samples were reduced on excel: On Peak Zero, sample-standard bracketing and blk correction.

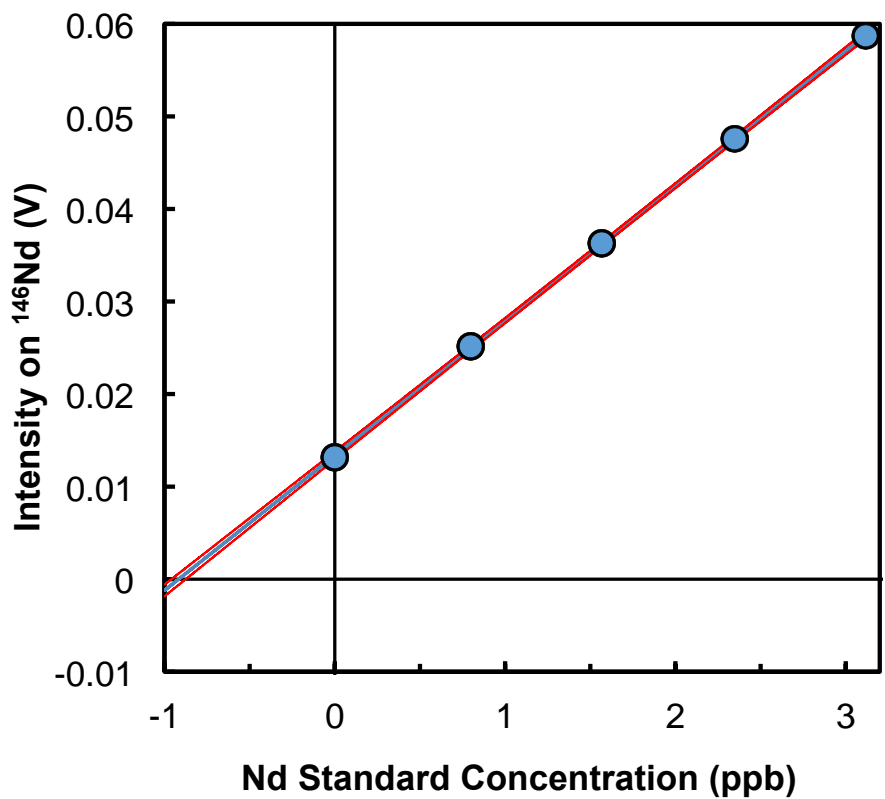

**fig. S17.** Results of the standard addition measurements conducted on the *Curious Marie* CAI. In standard addition, an array of solutions are prepared, all of which contain the same amount of sample solution (whose concentration is to be determined) and the same final volume, but varying amount of standard solution (of known concentration). In the above example, the relative proportion of sample/standard/Milli-Q water are (points from left to right): (1:0:4), (1:1:3), (1:2:2), (1:3:1) and (1:4:0). The sample concentration is obtained as the absolute value of the intercept of the regression (blue line) with the x-axis. The red curves show the 95 % confidence interval of the regression.

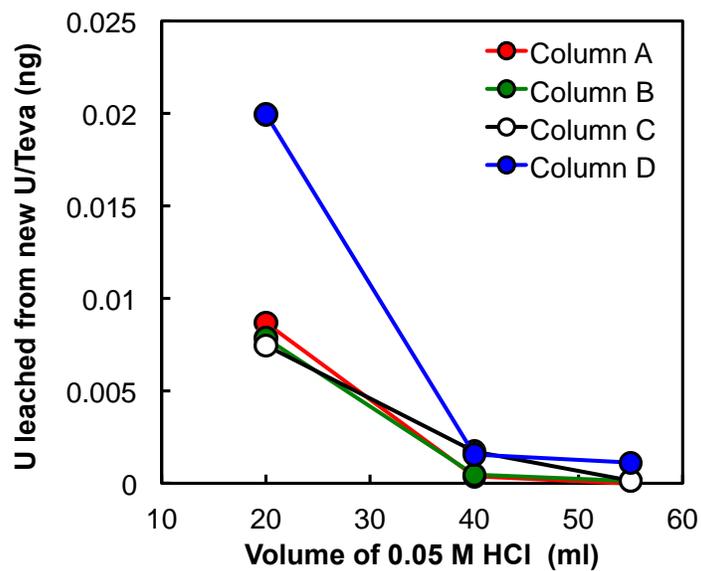

**fig. S18.** U blank from new U/Teva resin as a function of the volume of 0.05 M HCl passed through the column. The test was done on four new 2 mL pre-packed cartridge. After 40 ml of 0.05 M HCl, essentially all U initially bound to the resin is removed (down to the pg level). Proper cleaning on the resin is crucial when working on very low U content samples (*e.g.*, *Curious Marie*, replicate 2 and 3 contained only 0.048 and 0.032 ng of U, respectively).

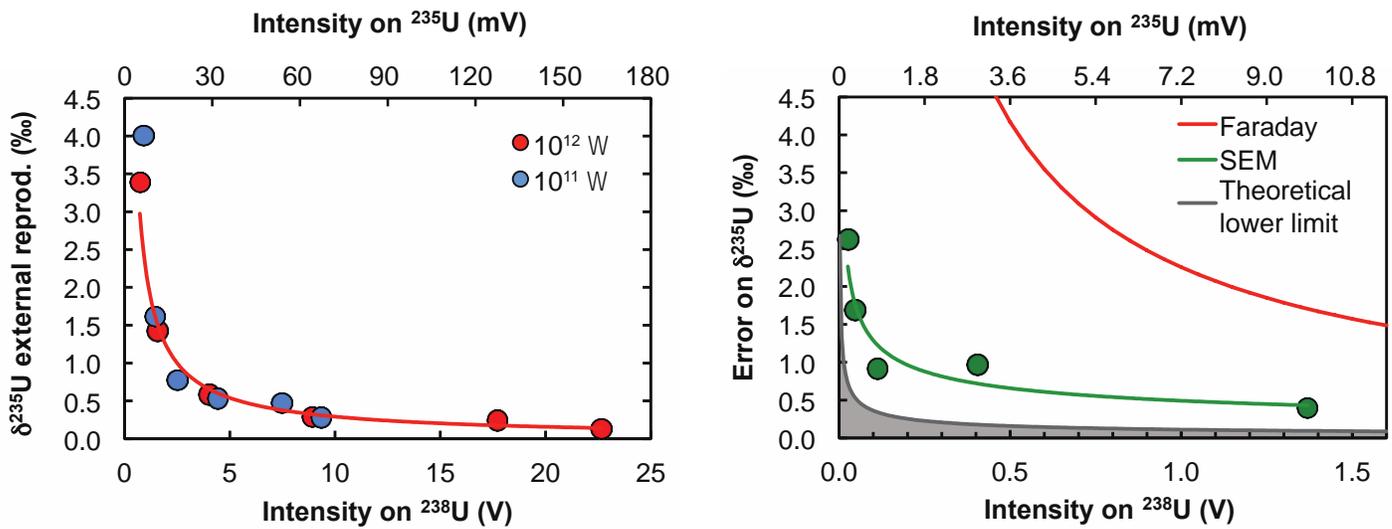

**fig. S19.** Results of precision tests done using two configurations: the *Faraday* setup, where $^{235}$U is measured on a Faraday cup (with either a $10^{12}$ $\Omega$ resistor or a $10^{11}$ $\Omega$ resistor) (left) and the *SEM* setup where $^{235}$U is measured on the SEM (right). In both cases $^{238}$U is being measured using a $10^{11}$ $\Omega$ resistor (see table S3 for detail of each setup). Each point represents 210 sec of analyses: 50 cycles of 4.194 sec for *Faraday* setup and 100 cycles of 2.097 sec for the *SEM* setup). The theoretical lower limit corresponds to the quadratic sum of the counting statistic and Johnson noise uncertainties, which were calculating using formulas (A.1) and (A.2) in (*58*). Note that the $^{235}$U intensity scale at the top of the graphs is the intensity reported by the MC-ICPMS software, which corresponds to the ion flux (in Ampere) normalized to a $10^{11}$ $\Omega$ resistor. For very low U concentrations, higher precision can be obtained by measuring $^{235}$U on the SEM rather than on a $10^{12}$ $\Omega$ or $10^{11}$ $\Omega$ resistor.

**Table S3**
Specifics of U isotopic measurements on MC-ICPMS for low U samples

| | L2 | L1 | Axial | H1 | H2 | H3 | Integration | Cycles | Rinse time |
|---|---|---|---|---|---|---|---|---|---|
| *Faraday* setup[a] | | | | | | | | | |
| Isotope | - | $^{233}$U | $^{234}$U | $^{235}$U | $^{236}$U | $^{238}$U | 2.097 sec | 100 | > 500 sec |
| Resistor (W) | - | $10^{11}$ | SEM | $10^{12}$ | $10^{11}$ | $10^{11}$ | | | |
| *SEM* setup[a] | | | | | | | | | |
| Isotope | $^{233}$U | $^{234}$U | $^{235}$U | $^{236}$U | $^{238}$U | - | 2.097 sec | 100 | > 500 sec |
| Resistor (W) | $10^{11}$ | $10^{11}$ | SEM | $10^{11}$ | $10^{11}$ | - | | | |

[a] For extremely small U amounts (0.01 to 0.05 ng of U) the *SEM* setup was used instead of the *Faraday* setup. For a comparison of achievable precisions with each setup see Fig. S19.

Table S4
Compilation of chemistry blanks and effect on U "stable" isotope ratio

| # | $H_2O_2$ | Session | Resins # | Cycle # | $^{235}U$ (V or cps) | ± (2SD) | $^{238}U$ (V) | ± (2SD) | $^{238}U/^{235}U$ blk | $d^{238}U$ blk (‰) | ± (2SD) | Sensitiviy (V/ppb) | Blk contribution (%)[a] | Effect on $d^{238}U$ meas. (‰)[a] |
|---|---|---|---|---|---|---|---|---|---|---|---|---|---|---|
| 1 | Yes | CAI Dec2014 a | 1 | 38 | 2.0E-04 | 6.1E-05 | 0.029 | 0.007 | 144.449 | 33 | 199 | 1.05 | 2.78% | 0.9 |
| 2 | Yes | CAI Dec2014 b | 1 | 90 | 1.1E-04 | 4.2E-05 | 0.015 | 0.004 | 138.208 | -12 | 315 | 1.05 | 1.41% | -0.2 |
| 3 | Yes | SEM Dec2014[b] | 2 | 39 | *4437.9* | *6417.8* | 0.010 | 0.015 | 133.409 | -46 | 83 | 1.44 | 0.66% | -0.3 |
| 4 | Yes | Far Jan 2015 | 1 | 33 | 2.2E-04 | 4.9E-05 | 0.035 | 0.005 | 157.504 | 126 | 136 | 1.58 | 2.24% | 2.8 |
| 5 | Yes | Far Mar 2015 | 3 | 24 | 6.2E-04 | 7.0E-05 | 0.086 | 0.006 | 137.740 | -15 | 98 | 1.42 | 6.07% | -0.9 |
| 6 | Yes | SEM Mar 2015[b] | 3 | 31 | *31877.0* | *2093.3* | 0.071 | 0.004 | 139.997 | 0.8 | 9.2 | 1.27 | 5.59% | 0.0 |

[a] The blank contribution and effect on the $d^{238}U$ measured are calculated assuming a sample solution at 1 ppb is measured (*i.e.*, about 1 to 1.5 V on $^{238}U$)
[b] The cps to volts conversion factor used in these sessions was estimated to be ~ 62e6 ± 0.3e6.

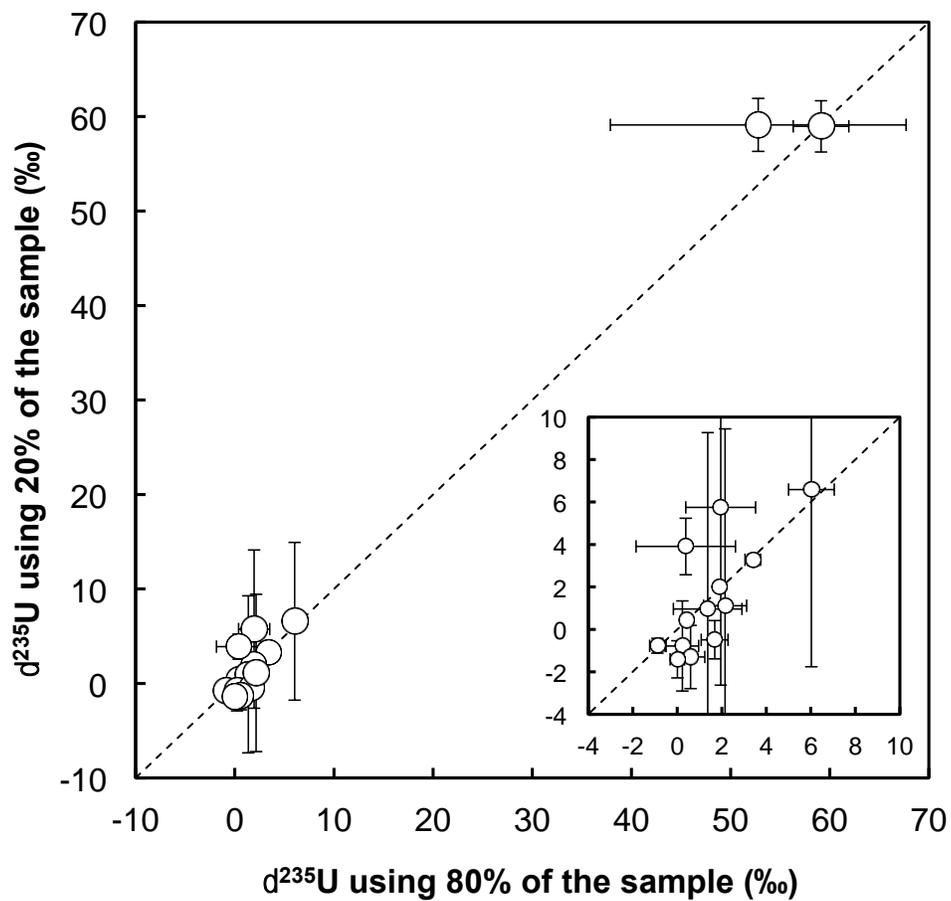

**fig. S20.** Comparison of the $\delta^{235}U$ determined using 80% (x-axis) and 20% (y-axis) of the sample. The *Curious Marie* CAI was measured three times (see fig. S21). All data are given relative to CRM-112a.

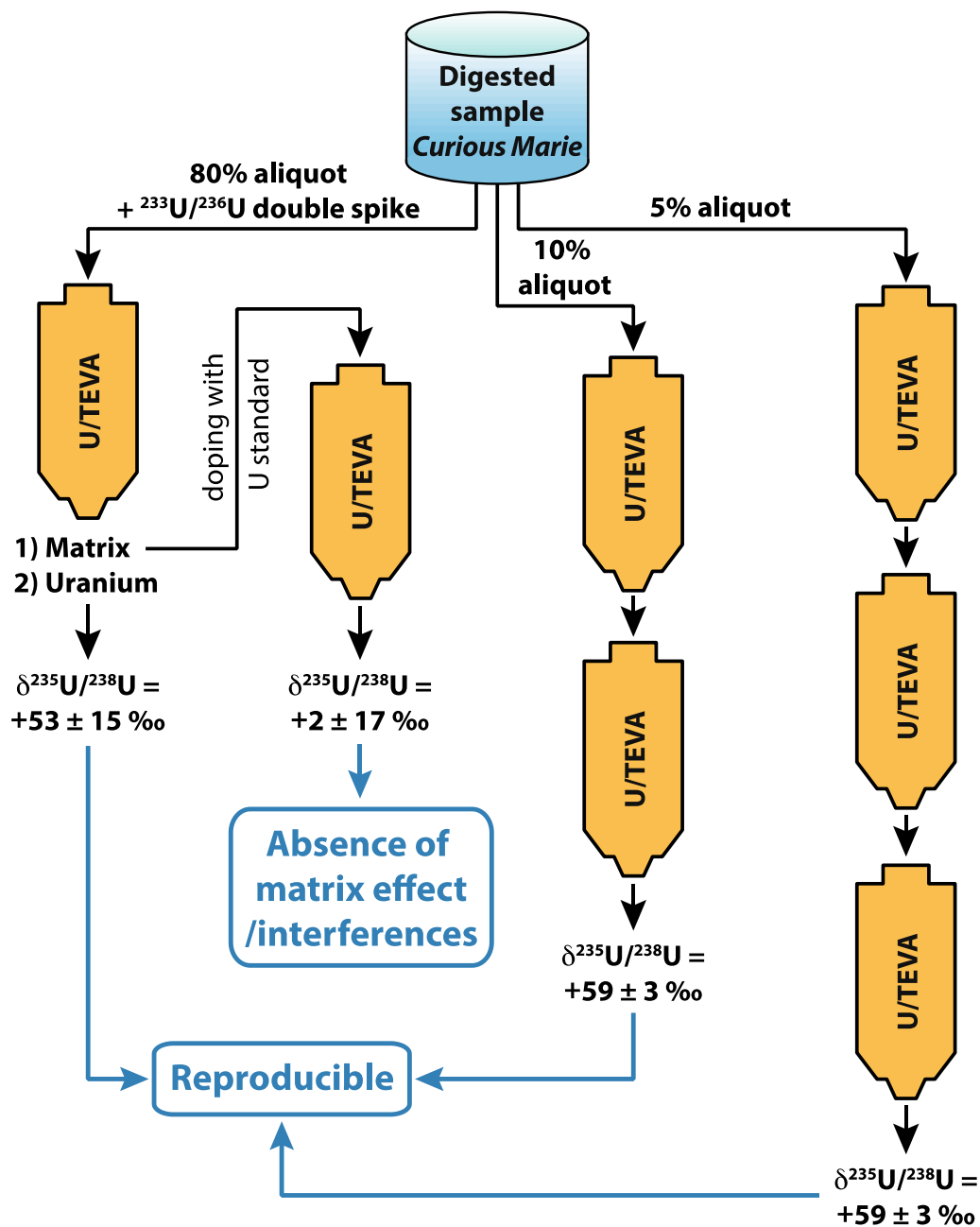

**fig. S21.** Flow chart of the tests conducted on the *Curious Marie* CAI. An 80% aliquot of the sample was spiked with IRMM-3636a ($^{233}$U/$^{236}$U double spike) and purified by column chemistry once. Analysis was done on a Neptune MC-ICPMS with $^{238}$U ($10^{11}$ Ω) and $^{235}$U ($10^{12}$ Ω) on faraday cups. The matrix of the sample (all elements but U) was doped with U standard CRM-112a, then subjected to one step of purification and analyzed to quantify potential matrix effects. The remaining 20% of the sample was split in two halves, purified two and three times, respectively. Analysis was done with $^{238}$U ($10^{11}$ Ω) on a faraday cup and $^{235}$U on the SEM. All measurements yielded identical values (within uncertainties).

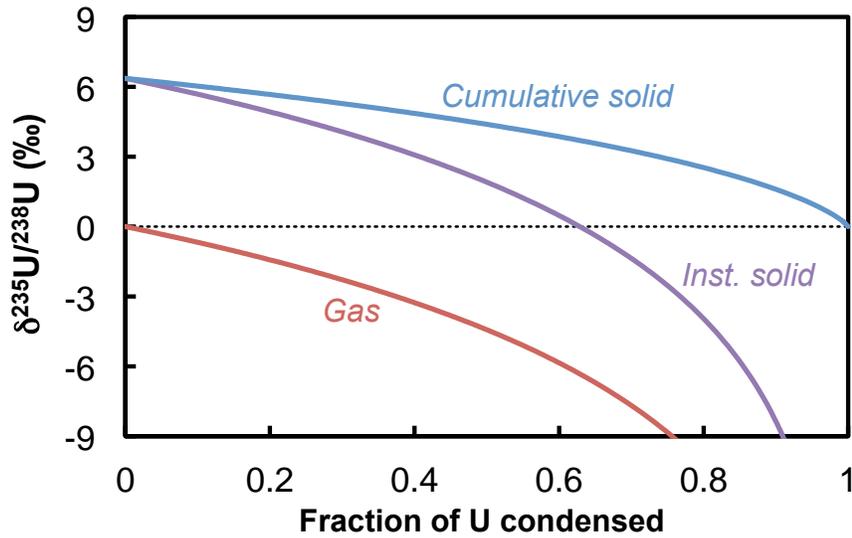

**fig. S22.** Evolution of the isotopic composition of the gas, instantaneous solid and cumulative solid, as a function of the fraction of U condensed. During condensation, the light isotope of U condenses faster, leading to large $^{235}$U-depletion in the condensing gas and the instantaneous solid. The maximum conceivable fractionation during condensation of U isotopes is given by the square root of the mass of the isotopes involved; $\alpha_{Cond} = \sqrt{238/235} = 1.0063$ [see (*59, 60*)]. This corresponds to a $^{235}$U-excesses of + 6.3‰. Similarly, during evaporation, the highest $^{235}$U-excess predicted by the kinetic theory of gases is ~6‰. Condensation/evaporation of molecular species, the presence of $H_2$, and evaporation/condensation in non-highly undersaturated or supersaturated conditions would all lead to smaller isotopic effects. Consequently, evaporation/condensation processes cannot explain the 59‰ $^{235}$U-excess observed in *Curious Marie*.

**Table S5**
Summary of the Ti data obtained on geostandards and the *Curious Marie* CAI.

| Samples | n | Mass-dependent fractionation + isotopic anomaly[a] | | | | | | | | | | | Isotopic anomalies | | | | | |
|---|---|---|---|---|---|---|---|---|---|---|---|---|---|---|---|---|---|---|
| | | Smp-std bracketing | | | | | | | | | | | Internal Normalization ($^{49}Ti/^{47}Ti$) | | | | | |
| | | $d^{46}Ti$ | ± (2s) | $d^{48}Ti$ | ± (2s) | $d^{49}Ti$ | ± (2s) | $d^{50}Ti$ | ± (2s) | $e^{46}Ti$ | ± (2s) | $e^{48}Ti$ | ± (2s) | $e^{49}Ti$ | ± (2s) | $e^{50}Ti$ | ± (2s) |
| BCR2-1 | 13 | -0.02 | 0.08 | 0.02 | 0.08 | 0.03 | 0.15 | 0.04 | 0.23 | -0.06 | 0.06 | 0.01 | 0.05 | 0 | 0 | -0.06 | 0.07 |
| BCR2-2 | 13 | -0.06 | 0.08 | 0.05 | 0.08 | 0.09 | 0.15 | 0.13 | 0.23 | -0.07 | 0.09 | 0.00 | 0.04 | 0 | 0 | -0.08 | 0.10 |
| *Curious Marie* | 4 | 0.43 | 0.18 | -0.23 | 0.13 | -0.55 | 0.30 | 0.07 | 0.47 | 1.46 | 0.44 | 0.48 | 0.27 | 0 | 0 | 8.90 | 0.26 |

[a] Values expressed in delta notation relative to the following natural abundances: n(46)/n(47)=1.09297, n(48)/n(47)=10.0690, n(49)/n(47)=0.74977, n(50)/n(47)=0.72917.
Errors are calculated as t*s/√n, where "t" is the student factor, "s" is the standard deviation of the replicate measurements, and "n" is the number of replicate measurements for a given sample solution.

**Table S6**
Production ratios of selected short-lived radionuclides produced by the s, r and p processes and present in the ESS, normalized to a stable isotope produced in the same or similar nucleosynthetic process.

| Nuclide | t = 1/λ (Myr) | Norm. | Meyer & Clayton 2000 | Jacobsen 2005 | Nittler and Dauphas 2006 | Huss et al., 2009 | Lugaro et al. 2014 | **Best estimate** | Refs.[a] |
|---|---|---|---|---|---|---|---|---|---|
| Niobium-92 | 50.1 | 92Mo | | | 0.0015 ± 0.0006 | | | **1.5E-03 ± 6E-04** | |
| Palladium-107 | 9.38 | 108Pd | 0.65 | 1.50 | | 1.29 | 2.09 | **1.37 ± 0.72** | * |
| Palladium-107 | 9.38 | 110Pd[r] | | | 1.36 | | | **1.36 ± 0.68** | *** |
| Iodine-129 | 23.1 | 127I | 1.40 | 2.00 | | 1.84 | | **1.70 ± 0.30** | * |
| Iodine-129 | 23.1 | 127I[r] | | | 1.45 | | 1.35 | **1.40 ± 0.70** | ** |
| Samarium-146 | 149 | 144Sm | 0.1 | 0.95 | 0.18 ± 0.06 | 3.05 | | **0.14 ± 0.07** | **,b,c |
| Hafnium-182 | 13 | 177Hf[r] | | | 0.81 | | | **0.55 ± 0.27** | [1] *** |
| Hafnium-182 | 13 | 180Hf | 0.21 | 0.37 | | 0.43 | 0.91 | **0.56 ± 0.35** | * |
| Plutonium-244 | 117 | 238U | | | 0.53 ± 0.36 | | | **0.53 ± 0.36** | [2] |
| Plutonium-244 | 117 | 232Th | 0.7 | 0.67 | 0.32 ± 0.27 | 1.33 | | **0.32 ± 0.27** | [2] |
| Curium-247 | 22.5 | 235U | | | 0.281 ± 0.165 | | 0.40 | **0.341 ± 0.170** | ** |
| Curium-247 | 22.5 | 238U | | | 0.159 ± 0.075 | | | **0.159 ± 0.075** | [2] |
| Curium-247 | 22.5 | 232Th | | | 0.098 ± 0.052 | | | **0.098 ± 0.052** | [2] |

The superscript r refers to the *r*-process component of the cosmic abundance, obtained after subtracting the *s*-process contribution from the solar abundances, from Bisterzo et al. (2014). Note that Nittler and Dauphas (2006) used the decomposition of Arlandini et al. (1999). The estimation of the r-component of 182W is significantly different in the two studies.

[a] References: [1] The best estimate of the 182Hf/177Hfr production ratio is calculated as the ratio of the *r*-component of 182W over 177Hf using the values from Bisterzo et al. (2014); [2] Goriely & Arnould, (2001); only the models giving a 232Th/238U production ratio consistent with the production ratio inferred from meteoritic measurements (1.75 +1.10/-1.01, Dauphas, 2005) are considered, (*i.e.*, cases 1, 3-4, 7-9, 13, 20, 22 and 30 in Table 1 and 2 of Goriely & Arnould (2001)).

[b] The value from Huss et al. (2009) is not used.

[c] The value from Jacobsen (2005) is not used.

*The best estimate is taken as the mid-range value from the four studies compiled in this table, with the errors reaching to the min. and max. values.

**The best estimate is taken as the average value from the four studies compiled in this table, with 50% relative uncertainty.

***Relative uncertainty assumed to be equal to 50%.

References used in this table:

Arlandini et al. (1999): (*61*)

Bisterzo et al. (2014): (*3*)

Dauphas (2005): (*62*)

Goriely & Arnould (2001): (*63*)

Huss et al. (2009): (*8*)

Jacobsen (2005): (*64*)

Nittler and Dauphas (2006): (*10*)

**Table S7**
Compilation of Cm/U isochron data and free decay interval data from experimental and theoretical studies.

| | Stirling et al. 2005 | Stirling et al. 2006 | Brennecka et al. 2010 | This work | Theoretical lower limit[c] |
|---|---|---|---|---|---|
| d$^{238}$U vs $^{144}$Nd/$^{238}$U intercept | 0.0072512 ± 1.9E-06 | 0.0072512 ± 1.2E-06 | 0.0072541 ± 7E-07 | 0.0072517 ± 1.2E-06 | |
| ($^{235}$U/$^{238}$U)$_{ESS}$ | 0.32111 ± 8.3E-05 | 0.32111 ± 5.5E-05 | 0.32124 ± 3.2E-05 | 0.32113 ± 5.2E-05 | |
| d$^{238}$U vs $^{144}$Nd/$^{238}$U slope | < 3.3E-08 | < 2.6E-08 | 3.7E-08 ± 7E-09 | 1.87E-08 ± 9.4E-10 | > 1.7E-08 ± 9E-09 |
| ($^{247}$Cm/$^{144}$Nd)$_{INITIAL}$ | | | | 1.65E-06 ± 8.29E-08 | |
| ($^{247}$Cm/$^{235}$U)$_{INITIAL}$ | | | | 5.57E-05 ± 2.8E-06 | |
| ($^{247}$Cm/$^{238}$U)$_{INITIAL}$ | | | | 1.79E-05 ± 9E-07 | |
| ($^{247}$Cm/$^{232}$Th)$_{INITIAL}$ | | | | 7.7E-06 ± 4E-07 | |
| Timing of alteration, (Myr after SS formation) | | | | 5 ± 5 | |
| ($^{247}$Cm/$^{144}$Nd)$_{ESS}$ | < 2.9E-06 | < 2.3E-06 | 3.3E-06 ± 6E-07 | 2.06E-06 ± 4.7E-07 | > 1.5E-06 ± 8E-07 |
| ($^{247}$Cm/$^{235}$U)$_{ESS}$ | < 9.8E-05 | < 7.7E-05 | 1.1E-04 ± 2E-05 | 7.0E-05 ± 1.6E-05 | > 5.0E-05 ± 2.6E-05 |
| ($^{247}$Cm/$^{238}$U)$_{ESS}$ | < 3.2E-05 | < 2.5E-05 | 3.5E-05 ± 6E-06 | 2.2E-05 ± 5E-06 | > 1.6E-05 ± 8E-06 |
| ($^{247}$Cm/$^{232}$Th)$_{ESS}$ | < 1.4E-05 | < 1.1E-05 | 1.5E-05 ± 3E-06 | 9.7E-06 ± 2.2E-06 | 6.9E-06 ± 3.6E-06 |
| *Using $^{238}$U as the normalizing isotope* | | | | | |
| R/P ($^{247}$Cm/$^{238}$U$_{stable}$) ESS[a] | < 1.4E-04 | < 1.1E-04 | 1.58E-04 ± 8.0E-05 | 1.00E-04 ± 5.2E-05 | 7.2E-05 ± 5.0E-05 |
| R/P ($^{247}$Cm/$^{238}$U$_{stable}$) ISM[b] | 7.0E-03 ± 1.6E-03 | | | | |
| **Free decay interval, D (Myr)** | **> 88** | **> 93** | **85 ± 12** | **95 ± 13** | **103 ± 17** |
| *Using $^{232}$Th as the normalizing isotope* | | | | | |
| R/P ($^{247}$Cm/$^{232}$Th$_{stable}$) ESS[a] | < 1.2E-04 | < 9.8E-05 | 1.4E-04 ± 7.8E-05 | 8.8E-05 ± 5.1E-05 | 6.3E-05 ± 4.7E-05 |
| R/P ($^{247}$Cm/$^{232}$Th$_{stable}$) ISM[b] | 7.0E-03 ± 1.6E-03 | | | | |
| **Free decay interval, D (Myr)** | **> 91** | **> 96** | **88 ± 14** | **98 ± 14** | **< 106 ± 17** |

The age of the solar system is taken as 4.567 Gyr, and the mean life (1/λ) of $^{247}$Cm = 22.5 Myr.
The present abundances of 144Nd (0.197), 232Th (0.0335), 235U (6.48e-5), and 238U (0.0089) are taken from Anders and Grevesse (1989).
The ESS abundances of 232Th (0.04197), 235U (0.00582), and 238U (0.0181) are calculated using λ232=4.93e-11, λ235=9.85e-10 and λ238=1.55e-10 yr$^{-1}$, respectively.
The production ratios of 247Cm/238U=0.159 ± 0.075 and 247Cm/232Th=0.098 ± 0.052 are taken from Goriely and Arnould (2001). Only the models giving a 232Th/238U production ratio consistent with the production ratio inferred from meteoritic measurements (1.75 +1.10/-1.01, Dauphas, 2005) are considered, (i.e., cases 1, 3-4, 7-9, 13, 20 ,22 and 30 in Table 1 and 2 of Goriely and Arnould (2001)).
[a] Because the isotopes used for nomarlization are not stable (only long-lived), the R/P ratio has to be corrected for the decay of the long-lived isotope. This is done by multiplying the R/P (247Cm/238U) ratio by the N/P ratio of 238U in the ESS (0.71) and the N/P (247Cm/232Th) ratio by the N/P ratio of 232Th in the ESS (0.89). Both values are taken from Nittler and Dauphas (2006).
[b] The N/P ratios in the ISM are calculated using Eq. (2) of the main text (open nonlinear GCE model of Dauphas et al. (2003)), with a k=1.7±0.4 (2s), and a presolar age of the galaxy Tg=8.7±1.5 Gyr (2s).
[c] The lower limit on the (247Cm/235U)ESS ratio from Nittler and Dauphas (2006) is used as input value for this column.

References used in this table:

Anders and Grevesse (1989): (*65*)

Dauphas et al. (2003): (*28*)

Dauphas (2005): (*62*)

Goriely & Arnould (2001): (*63*)

Nittler and Dauphas (2006): (*10*)

**Table S8**
Selected extinct radionuclides produced by the s, r and p processes.

| Nuclide | | $t = 1/\lambda$ (Myr) | Norm. | R | Refs | P | R/P[a] |
|---|---|---|---|---|---|---|---|
| Niobium-92 | p-process | 50.1 | 92Mo | $(2.8 \pm 0.5) \times 10^{-6}$ | [1] | $(1.5 \pm 0.6) \times 10^{-3}$ | $(1.9 \pm 0.8) \times 10^{-2}$ |
| Palladium-107 | s+r process | 9.38 | 108Pd | $(5.9 \pm 2.2) \times 10^{-5}$ | [2] | $1.4 \pm 0.7$ | $(4.3 \pm 2.8) \times 10^{-5}$ |
| | | | 110Pd[r] | $(1.4 \pm 0.5) \times 10^{-4}$ | [2] | $1.4 \pm 0.7$ | $(1.0 \pm 0.6) \times 10^{-4}$ |
| Iodine-129 | r-process | 23.1 | 127I | $(1.2 \pm 0.2) \times 10^{-4}$ | [3] | $1.7 \pm 0.3$ | $(7.0 \pm 1.7) \times 10^{-5}$ |
| | | | 127I[r] | $(1.3 \pm 0.2) \times 10^{-4}$ | [3] | $1.4 \pm 0.7$ | $(8.9 \pm 4.7) \times 10^{-5}$ |
| Samarium-146 | p-process | 149 | 144Sm | $(8.4 \pm 0.5) \times 10^{-3}$ | [4] | $(1.4 \pm 0.7) \times 10^{-1}$ | $(6.0 \pm 3.0) \times 10^{-2}$ |
| Hafnium-182 | s+r-process | 13 | 177Hf[r] | $(2.2 \pm 0.1) \times 10^{-4}$ | [5] | $(5.5 \pm 2.8) \times 10^{-1}$ | $(4.1 \pm 2.1) \times 10^{-4}$ |
| | | | 180Hf | $(9.7 \pm 0.4) \times 10^{-5}$ | [5] | $(5.6 \pm 3.5) \times 10^{-1}$ | $(1.7 \pm 1.1) \times 10^{-4}$ |
| Plutonium-244 | r-process | 117 | 232Th | $(2.9 \pm 0.4) \times 10^{-3}$ | [6] | $(3.2 \pm 2.7) \times 10^{-1}$ | $(8.1 \pm 7.1) \times 10^{-3}$ |
| Curium-247 | r-process | 22.5 | 232Th | $(9.7 \pm 2.2) \times 10^{-6}$ | [7] | $(9.8 \pm 5.2) \times 10^{-2}$ | $(8.8 \pm 5.1) \times 10^{-5}$ |

The superscript r refers to the r-process component of the cosmic abundance [obtained after subtracting the s-process contribution from solar abundances, from Arlandini et al. (1999); Bisterzo et al. (2014)]. R is the initial abundance ratio in the ESS, P is the production ratio (see best estimate values in Table S6).
References: [1] Harper 1996; Schönbächler et al., 2002; [2] Chen & Wasserburg 1996; Schönbächler et al., 2008; [3] Brazzle et al., 1999; Reynolds 1960; [4] Boyet et al. 2010; Lugmair & Galer 1992; Prinzhofer et al., 1992; [5] Burkhardt et al., 2008; Kleine et al., 2002; Yin et al., 2002; [6] Hudson et al., 1989; [7] This work; Brennecka et al., 2010.
[a] When the isotope used for normalization is not stable (e.g., X/238U or X/232Th), the R/P ratio has to be corrected for the decay of the long-lived isotope. This is done by multiplying the R/P ratio by the N/P ratio of the normalizing isotope in the ESS (0.71 for 238U and 0.89 for 232Th, both values are taken from Nittler and Dauphas (2006)).

References used in this table:

Arlandini et al. (1999): (*61*)
Bisterzo et al. (2014): (*3*)
Boyet et al. (2010): (*66*)
Brazzle et al. (1999): (*67*)
Brennecka et al. (2010): (*11*)
Burkhardt et al. (2008): (*68*)
Chen & Wasserburg (1996): (*69*)
Harper (1996): (*70*)
Hudson et al. (1989): (*71*)
Kleine et al. (2002): (*72*)
Lugmair & Galer (1992): (*73*)
Nittler and Dauphas (2006): (*10*)
Prinzhofer et al. (1992): (*74*)
Reynolds (1960): (*75*)
Schönbächler et al. (2002): (*76*)
Schönbächler et al. (2008): (*77*)
Yin et al. (2002): (*78*)